\documentclass[12pt]{article}
\usepackage{amsmath,amsfonts,amssymb}
\makeatletter \@addtoreset{equation}{section}

\makeatletter\renewcommand\section{\@startsection {section}{1}{\z@}%
                                   {-3.5ex \@plus -1ex \@minus -.2ex}
                                   {2.3ex \@plus.2ex}%
                                   {\normalfont\large\bfseries}}
\renewcommand\subsection{\@startsection{subsection}{2}{\z@}%
                                     {-3.25ex\@plus -1ex \@minus -.2ex}%
                                     {1.5ex \@plus .2ex}%
                                     {\normalfont\bfseries}}

\newcommand{\be}{\begin{equation}}
\newcommand{\ee}{\end{equation}}
\newcommand{\beq}{\begin{eqnarray}}
\newcommand{\eeq}{\end{eqnarray}}

\newcommand{\bbibitem}[1]{\bibitem{#1}\marginpar{#1}}

\def\Label#1{\label{#1}%
  \smash{\hbox to0pt{\raise1ex\hbox{\tiny[#1]}\hss}}}
\def\noLabels{\let\Label=\label}
\def\nobbibitem{\let\bbibitem=\bibitem}

\newcommand{\G}{\Gamma}

\def\S{\Sigma}

\def\CA{{\cal A}}

\def\CN{{\cal N}}

\newcommand{\SL}{\mathrm{SL}}

\newcommand{\RR}{\mathbb{R}}

\begin{document}

\begin{titlepage}

\vfil\

\begin{center}

{\Large{\bf Deconstructing the D0-D6 system\\ }}

\vspace{3mm}

Alejandra Castro\footnote{e-mail: aycastro@umich.edu}$^{\,a}$ and
Joan Sim\'on\footnote{e-mail: J.Simon@ed.ac.uk}$^{\,b}$
\\

\vspace{8mm}

\bigskip\medskip
\smallskip\centerline{$^a$ \it Department of Physics, University of Michigan, Ann Arbor,
MI 48109, USA.}
\medskip
\smallskip\centerline{$^b$ \it School of Mathematics and Maxwell Institute for Mathematical Sciences,}
\smallskip\centerline{\it King's Buildings, Edinburgh EH9 3JZ, Scotland.}

\vfil

\end{center}
\setcounter{footnote}{0}
\begin{abstract}
\noindent We find the complete classical moduli space of two-centered
supersymmetric solutions carrying D0 and D6 brane charge in the STU
model delimited by walls of marginal stability of co-dimension one.
U-duality guarantees our conclusions hold for any BPS state with
negative quartic invariant. The analysis explicitly shows that the
conditions of marginal stability, {\it i.e.} the integrability
conditions, are generically insufficient to provide a regular
supergravity solution in this model.

\end{abstract}
\vspace{0.5in}

\end{titlepage}
\renewcommand{\baselinestretch}{1.05}  

\newpage
\tableofcontents

\newpage
\section{Introduction}

String theory provides a microscopic description of black holes as
bound states of D-branes and other solitonic objects. Significant
progress has been achieved by understanding the structure of these
bound states and how these features manifest in supergravity
\cite{Douglas:1999vm,Denef:2000ar,Denef:2001xn,Denef:2002ru,Sen:2007vb,Dabholkar:2007vk,Cheng:2007ch,Sen:2007qy,Cheng:2008fc}.
In asymptotically flat four dimensional spacetimes, some BPS states
in string theory with a fixed set of charges can be described as a
single center black hole and/or as a multi-centered solution
\cite{Behrndt:1997ny,Denef:2000nb,LopesCardoso:2000qm,Bates:2003vx,Sen:2007pg}.
Generically, the asymptotic conserved charges and a set of regularity
conditions define a classical moduli space, which should be after
proper quantization in agreement with the microscopic theory in
appropriate regimes \cite{Grant:2005qc,Maoz:2005nk,deBoer:2008zn}.
The split attractor flow conjecture \cite{Denef:2001xn,Denef:2007vg}
proposes a description of this moduli space for half BPS states in
${\cal N}=2$ theories in 4D. The basic idea is that a solution will
exist if there is an attractor flow tree in moduli space that
terminates on the attractor points of the constituents charges. The
bifurcation points of the tree correspond to the regions in moduli
space where the state becomes marginally stable and breaks apart.

For a given bound state with fixed charge vector, {\it a priori}
there may be an infinite number of ways to split up its charge into
bound state composites. This would lead to an infinite degeneracy
which is known not to occur. As conjectured by the split attractor
flow, there should be physical requirements that will only allow a
finite number of such decompositions. In supergravity, these
translate to kinematic conditions ({\it e.g.} mass and charge
conservation) and dynamical conditions ({\it e.g.} smooth geometry)
on the multi-centered solution describing the bound state. Our
motivation is to investigate these conditions in detail and classify
all possible composites for a given total charge.

Answering this question is extremely difficult for a generic
supersymmetric bound state. In this paper, we will focus on a
particular class of BPS states with negative quartic invariant,
$\Delta$, and study the realization of supersymmetric states in
the STU model. States with $\Delta<0$ are particulary interesting
because they will always correspond to polar states in the BPS
branch, {\it i.e.} the supergravity description is always
multi-centered. In addition, U-duality guarantees that we can choose
a U-dual frame where the system only carries D0 and D6-brane charges
\cite{ferrara-latest}. This is an extremely simple charge vector
which will allow us to explicitly construct the bound states in a
fairly straight forward manner.


The D0-D6 system was analyzed in
\cite{Kachru:1999vj,Mihailescu:2000dn,Douglas:2000ah,Witten:2000mf,Fujii:2001wp},
where the existence of supersymmetric bound state was guaranteed if a
sufficiently large $B$-field was turned on. This condition defines a
region of moduli space where the state exists, and it is delimited by
a wall of marginal stability of co-dimension one. More recently,
these bound states have been described in the large volume
approximation as two-centered supergravity configurations
\cite{Denef:2007vg}, where one center carries D0-charge and the
second one carries D6-charge. The location in moduli space where the
bound state starts to exist in the classical theory coincides with
the wall of marginal stability derived in the weakly coupled
description of the D-brane system.


In the supergravity approximation, it is natural to ask whether there
are any other supersymmetric two-centered regular configurations
carrying the same charge as a D0-D6 bound state, but with different
constituent charges.\footnote{In some recent papers
\cite{Lee:2008ha,Camps:2008hb,Gimon:2009gk} similar questions have
been discussed for both the BPS and non-BPS branch of the D0-D6
system.} In the following we will determine all such configurations
that are bounded by co-dimension one walls of marginal stability in
moduli space. In principle one could consider solutions with more
than two centers, but the integrability conditions will generate
walls of higher co-dimension.

Our strategy will consist of two main steps: an algebraic
classification of the potential composites of the bound state and the
supergravity description of the latter. In the first step, we will
determine all possible candidate constituents building a D0-D6 bound
state, consistent with supersymmetry, and conservation of mass and
charge. Knowing the composite charges and fixing the moduli at
infinity, we can compute the central charges (in the large volume
limit) associated with these states and study the regions in moduli
space where they remain finite. Furthermore, we can also determine the
loci in moduli space where walls of marginal stability exist.

In the second step, we will find the supergravity realization for
these bound states as two-centered configurations and study their
regularity. We fix both the charges at infinity and at each center
(using the results in the first part of our analysis), and determine
the distance scale between the centers by solving the integrability
condition. This is guaranteed to be positive in the same region
defined by the wall of marginal stability, but it is not enough to
assure the regularity of the supergravity configuration. This requires, in addition,
the positivity of an scalar function $\Sigma^2$ and the absence of closed timelike
curves (CTCs). We will explicitly see that these requirements are
non-trivial. In particular, we will prove that all the conditions
required on the central charges in the first part of our analysis are
necessary, but still not sufficient to guarantee the existence of the
bound state in supergravity.

One main lesson of our analysis is to explicitly show that the
kinematic conditions derived from supersymmetry, in addition to
having well-defined composite states, are not enough to assure the
stability of the bound state. There are some non-trivial dynamical
conditions which in supergravity arise from requiring a regular
geometry. It would be interesting to understand how these conditions
are translated on the microscopic Hilbert space of BPS states.

This paper is organized as follows. In section 2, we start by briefly
reviewing the STU model and its most general stationary BPS
solutions. We comment on the connection between the zeroes in the
central charge and the location of the walls of marginal stability.
We also review how U-duality orbits allow us to focus on the D0-D6
system. In section 3, we first determine all 1/4 and 1/2 BPS charge
vectors consistent with conservation of mass and charge. We analyze
the conditions under which their central charges do not vanish and
determine the equations describing the walls of marginal stability in
each case. In section 4, we study the regularity of the corresponding
two-centered supergravity configurations. In section 5, we extend our
analysis to include 1/8 constituent BPS states and we finish with
some conclusions.

\section{D0-D6 in the STU model}

\subsection{STU model}

We begin the discussion with a brief overview of four dimensional BPS
configurations in supergravity. Our focus is on the $\CN=2$ theory
known as the STU-model \cite{Cremmer:1984hj,STU,Behrndt:1996hu}. We
will interpret the model in terms of type IIA string theory compactified on a $T^6$ of the
form $T^2\times T^2\times T^2$. The D0/D2/D4/D6-branes wrapping the
various cycles of $T^6$ give rise to four magnetic and four electric
charges that are assembled into the charge vector
\begin{equation}
  \Gamma = \left(\,p^0\,,\,p^A\,;\,q_A\,,\,q_0\,\right)\,,
 \label{eq:chvector}
\end{equation}
with $A=1,2,3$, and each component representing (D6,D4,D2,D0) brane
charges respectively. $\CN=2$ theories are characterized by a
prepotential $F$. In the STU model the prepotential and its
derivatives are
\be F = -\frac{X^1X^2X^3}{X^0}~,\qquad F_\Sigma = {\partial F\over
\partial X^\Sigma}~. \ee
We gauge fix the projective coordinates $X^\Lambda$
($\Lambda=0,1,2,3$) so that $X^0=1$, and define $X^A\equiv z^A =
B^A+iJ^A$. Then the K\"{a}hler potential is given by
\be \label{Kahler} K = - \ln i(F_\Sigma \bar
X^\Sigma -\bar F_\Sigma X^\Sigma) = - \ln( 8 J^1J^2J^3)~,
\ee
where $X^\Lambda\bar F_\Lambda = -X^0\bar F_0 + X^A\bar F_A$, and the central charge reads
\begin{equation}
  Z =  e^{K/2}\,[X^\Lambda q_\Lambda - F_\Lambda p^\Lambda] =
   e^{K/2}\,[ p^0\,z^1\,z^2\,z^3 -\frac{1}{2}s_{ABC}p^A\,z^B\,z^C + z^A\,q_A  -q_0] ~,
 \label{eq:ccharge}
\end{equation}
where the only non-vanishing intersection numbers are $s_{123}=1$ and
cyclic permutations.

\subsubsection{BPS solutions}\label{sec:BPSsol}

The most general stationary but non-static BPS configurations solving
the STU equations of motion were constructed in
\cite{Denef:2000nb,LopesCardoso:2000qm,Bates:2003vx} and are reviewed
in appendix \ref{sec:sugra}. Their metrics
\begin{equation}
  ds^2 =  -\frac{1}{\S}(dt+\omega)^2+\S\, ds_{\mathbb{R}^3}^2\,,
\end{equation}
are described by the one-form $\omega$ defined on $\mathbb{R}^3$ and
the scalar function $\Sigma^2$
\begin{align}\label{sigma}
  \Sigma^2(H) = &-(H_\Lambda H^\Lambda)^2+4\left(H^1H_1H^2H_2+H^1H_1H^3H_3+H^2H_2H^3H_3\right)\nonumber\\ &-4H^0H_1H_2H_3-4H_0H^1H^2H^3~,
\end{align}
depending on eight harmonic functions $(H^\Lambda,\,H_\Lambda)$
\begin{equation}\label{harmonic}
H^\Lambda=\sum_{i=1}^N {p^\Lambda_i\over |\vec{x}-\vec{x}_i|}+h^\Lambda~,\quad
H_\Lambda=\sum_{i=1}^N {q^i_\Lambda\over |\vec{x}-\vec{x}_i|}+h_\Lambda~.
\end{equation}
These harmonic functions encode all the information about the
conserved charges and moduli. The total charge
$\Gamma=(p^\Lambda\,;q_\Lambda)$ is split into $N$ centers, each
carrying charge vector $\Gamma_i = (p^\Lambda_i\,; q^i_\Lambda)$ so
that $p^\Lambda=\sum_i p^\Lambda_i$ and $q_\Lambda=\sum_i
q^i_\Lambda$. The moduli values at infinity $(z^A_\infty)$ and the
charge vector $\Gamma$ define a total central charge $Z=
|Z|\,e^{i\,\alpha}$. These determine the set of constants
$h=(h^\Lambda;\,h_\Lambda)$ (see \eqref{moduli}) by requiring the
metric to be asymptotically flat and to solve the integrability
conditions below.

Such solutions are regular if they satisfy:
\begin{itemize}
\item[1.] integrability conditions which guarantee the absence
    of Dirac-Misner strings
\begin{equation}\label{intmult}
  \sum_{b \neq a} \frac{\langle
    \G_a,\G_b\rangle}{r_{ab}}=\langle h,\G_a\rangle\,, \quad \text{with} \quad \langle \G_i,\,\G_j\rangle = -p_i^0q_0^j + p_i^Aq_A^j - q_A^ip^A_j + q_0^ip_j^0\,,
\end{equation}
\item[2.] positivity of the function $\Sigma^2$, {\it i.e.} $\Sigma^2>
    0 \,\,\,\forall\,\vec{x}\in \RR^3$,
\item[3.] absence of CTCs, {\it i.e.} $\Sigma^2-\omega_i\omega^i> 0  \,\,\,\forall\,\vec{x}\in \RR^3$, and absence of singularities in the moduli fields.
\end{itemize}

Close to each pole $\vec{x}_i$, the attractor equations govern the
behavior of the function $\Sigma^2$ and fixes the scalar moduli
\cite{Ferrara:1995ih,Strominger:1996kf,Ferrara:1996dd}. In
particular, the leading term as $\vec{x}\to\vec{x}_i$ is
\begin{equation}
\Sigma^2(\vec{x}\to\vec{x}_i) =  \frac{\Delta_i}{|\vec{x}-\vec{x}_i|^4} + {\cal O}\left({|\vec{x}-\vec{x}_i|^{-3}}\right)~,
\end{equation}
where $\Delta_i$ is  the quartic invariant associated to the charge
vector $\Gamma_i$. In the STU model, the quartic invariant of the
U-duality group $\left(\SL(2,\RR)\right)^3$ is given by
\begin{equation}\label{quartic}
\Delta=-(p^\Lambda q_\Lambda)^2+4\left(p^1q_1p^2q_2+p^1q_1p^3q_3+p^2q_2p^3q_3\right)-4p^0q_1q_2q_3-4q_0p^1p^2p^3
\end{equation}
with
\begin{equation*}
p^\Lambda q_\Lambda\equiv-p^0q_0+p^1q_1+p^2q_2+p^3q_3~.
\end{equation*}
The value of $\Delta$ determines the amount of supersymmetry
preserved by the system \cite{ferraramald}. For $\Delta>0$ we have a BPS black hole
preserving 1/8 supercharges; single centered solutions with $\Delta<0$ are non-BPS;
and if $\Delta=0$ the system can preserve 1/8 or more supercharges. As reviewed in \eqref{susy}
different BPS states have different scaling in $|\vec{x}-\vec{x}_i|$ \cite{4dmicro}.

\paragraph{Two-centered solutions.} The bound states we will construct in the later
sections consist on only two centers, hence it will be useful to
simplify the above expressions for such case. We will use a similar
notation to the one discussed in \cite{Lee:2008ha}. For any
two-centered configuration, we can always take the first center at
the origin and the second on the $z$-axis at distance $R$, carrying
generic charge vectors
\begin{align}
\Gamma_1&=(p^\Lambda_1,q_{\Lambda}^1)~,\quad \vec{x}_1=(0,0,0)~,\\
\Gamma_2&=(p^\Lambda_2,q_{\Lambda}^2)~,\quad \vec{x}_2=(0,0,R)~,
\end{align}
with $\langle \Gamma_1,\Gamma_2\rangle\neq0$. The harmonic functions
are given by \eqref{harmonic}, and by using standard spherical
coordinates on $\RR^3$ their radial dependence simplifies to
\begin{align}\label{polar}
|\vec{x}-\vec{x}_1|^2=r^2~, \quad \Theta^2\equiv|\vec{x}-\vec{x}_2|^2=r^2-2rR\cos\theta+R^2~.
\end{align}
The integrability conditions \eqref{intmult} are
\begin{align}\label{int}
{\langle \Gamma_1,\Gamma_2\rangle\over R}&={{\rm Im}(Z_1\bar{Z}_2)\over 2|Z_{1+2}|}\nonumber\\
&= h^\Lambda q^1_\Lambda - h_\Lambda p^\Lambda_1 = -h^\Lambda q^2_\Lambda + h_\Lambda p^\Lambda_2~.
\end{align}
Next, the one-form is determined by integrating \eqref{omegaeom}.
Using \eqref{int}, the right hand side of \eqref{omegaeom} reads
\begin{align}
\langle dH,H\rangle =-{\langle \Gamma_1,\Gamma_2\rangle\over R}\left(dr^{-1}-d\theta^{-1}\right)+
{\langle \Gamma_1,\Gamma_2\rangle}\left(\Theta^{-1}dr^{-1}-r^{-1}d\theta^{-1}\right)
\end{align}
Integrating the above expression, we obtain
\begin{align}\label{omega}
\omega={\langle \Gamma_1,\Gamma_2\rangle\over R}\left[1-{r+R\over \Theta}\right](1-\cos\theta)d\phi~,
\end{align}
where we fixed the integration constant so that our solutions are
asympotically flat, {\it i.e.} $\omega\to 0$ at infinity, and it
avoids Dirac-Misner singularities at $\theta=0,\pi$. Knowing
$\Sigma^2$ and the one-form $\omega$, the sufficient condition to
ensure the absence of CTCs is
\begin{equation}\label{eq:ctc}
  \Sigma^2\,r^2\,\sin^2\theta > \left(\omega_\phi\right)^2\,.
\end{equation}

\subsection{D0-D6 bound states}

We want to identify the possible different representations of the
D0-D6 system as a BPS bound state in the STU model. In the notation
introduced above, D0-D6 corresponds to turning only $p^0$ and $q_0$
in \eqref{eq:chvector}. The quartic invariant \eqref{quartic} is then
given by $\Delta=-(p^0q_0)^2$. Since the value of the $\Delta$ is
negative, it is clear that $\Sigma^2$ is not positive definite. In
particular, close to the charge source location $(\vec{x}\to 0)$,
\begin{equation*}
  \Sigma^2 \to - \left(\frac{p^0\,q_0}{|\vec{x}|^2}\right)^2 + {\cal O}\left({|\vec{x}|^{-3}}\right)~.
\end{equation*}
This observation is consistent with the existence of loci in moduli
space where the total D0-D6 central charge vanishes
\begin{equation*}
  Z_{D0-D6} = e^{K/2}\,\left(p^0\,z^1\,z^2\,z^3 - q_0\right) = 0 \quad \Leftrightarrow \quad {\rm Im}(z^1\,z^2\,z^3)=0\,,\,\,\, \,{\rm Re}(z^1\,z^2\,z^3) = {q_0\over p^0}
\end{equation*}
Whenever this occurs at a finite point in moduli space, the BPS state
does not exist, as argued in \cite{Seiberg:1994rs,
Strominger:1995cz,Moore:1998zu,Moore:1998pn}. We will use this
criterion all along this work.

The above conclusion was reached in the supergravity approximation
and assuming the realization of the state in terms of a single center
configuration. But D0-D6 states may allow different descriptions as a
function of the string coupling constant. The problem of adhering
D0-branes to D6-branes in a supersymmetric manner was studied in
\cite{Witten:2000mf}.\footnote{See also
\cite{Douglas:2000ah,Mihailescu:2000dn,Kachru:1999vj,Fujii:2001wp}}
It was found that a supersymmetric branch exists for sufficiently
large $B$-fields such that
\begin{equation}
\label{eq:boundary}
\frac{1}{2}s_{ABC}B^A\,J^B\,B^C \geq J^1\,J^2\,J^3~.
\end{equation}

In recent work in the supergravity literature \cite{Denef:2000nb},
these supersymmetric bound states were identified with two-centered
supergravity configurations carrying D6-brane and D0-brane charges at
each center. These are characterised by two charge vectors
\begin{equation*}
  \Gamma_1 = (p^0,\vec{0}; \vec{0},0) \quad \text{and} \quad \Gamma_2 = (0,\,\vec{0}; \vec{0}, q_0)\,,
\end{equation*}
sourced at points $\vec{x}_1$ and $\vec{x}_2$ separated by a distance
$R=|\vec{x}_1-\vec{x}_2|$, which is uniquely determined by solving
the integrability condition
\begin{equation}
  R={{\langle \Gamma_1,\Gamma_2\rangle}|Z_{1+2}| \over 2\,{\rm Im}(Z_1\bar{Z}_2) }~.
\end{equation}
The separation becomes infinite precisely when the equality in
\eqref{eq:boundary} is saturated, which corresponds to the location
of a wall of marginal stability
\begin{equation*}
  {\rm Im}(Z_1\bar{Z}_2) = 0\,.
\end{equation*}
This is interpreted as the disappearance of the bound state when
crossing such wall. Thus, for the bound state to exist the separation
scale $R$ must be physical, {\it i.e.} $\langle
\Gamma_1,\Gamma_2\rangle\,{\rm Im}(Z_1\bar{Z}_2) > 0$ is a necessary
condition. This can be confirmed by computing the number of BPS
states as a function of the moduli and proving the existence of a
jump in the mathematical index that accounts for these degeneracies
\cite{Denef:2007vg,Sen:2007pg,Cheng:2007ch}. \footnote{The agreement
of the BPS moduli space for the supergravity solution and the open
string perturbative analysis was explained in \cite{Denef:2000nb} and
compared to its non-BPS branch in \cite{Gimon:2009gk}.}

\subsection{Walls of marginal stability: systematics}
\label{sec:stability}

Given the connection between the existence of a D0-D6 BPS bound state
and a two-centered supergravity configuration, it is natural to
wonder whether there could be other two-centered configurations with
the same charges at infinity but different charge split
decomposition, {\it i.e.} different pole charge vectors
$\{\Gamma_1,\,\Gamma_2\}$.\footnote{In this work, we will focus on
the STU truncation of the full ${\cal N}=8$ supergravity, and the
reader should be aware that our conclusions may not apply to the full
theory.} This requires us to identify the different walls of marginal
stability where the split may occur. Given a BPS state with charge
vector $\Gamma_{1+2}$, mass $M_{1+2}=|Z_{1+2}|$ and central charge
$Z_{1+2}=e^{i\alpha}|Z_{1+2}|$, the necessary conditions that define
a wall of marginal stability are
\begin{subequations}\label{cd}
\begin{align}
\Gamma_{1+2}&=\Gamma_1+\Gamma_2\\
|Z_{1+2}|&=|Z_1|+|Z_2|
\end{align}
\end{subequations}
where $\{Z_i,\,\Gamma_i\}$ with $i=1,2$ stand for the data of the
bound state constituents once the wall is crossed. These conditions
assure conservation of charge and mass at the wall of marginal
stability. They are equivalent to solving
\begin{equation}
\begin{array}{cc}
  \text{Im}\left(Z_1\,\bar{Z}_2\right) &= 0\,,  \label{eq:noim} \\
  \text{Re}\left(Z_1\,\bar{Z}_2\right) &> 0\,.
\end{array}
\end{equation}
%


\paragraph{D0-D6 walls.} Given a D0-D6 system where
\begin{equation*}
  \Gamma_{\text{D0-D6}} = (p^0,\,\vec{0};\,\vec{0},\,q_0)\,,
\end{equation*}
its most general split into two vectors $\Gamma_1$ and $\Gamma_2$,
consistent with charge conservation, is
\begin{align*}
  \Gamma_1 &= \left(P^0,\,P^A;\,Q_A,\,Q_0\right)\,, \\
  \Gamma_2 &= \left(p^0-P^0,\,-P^A;\,-Q_A,\,q_0-Q_0\right)\,.
\end{align*}
The central charges of all the above charge vectors are
\begin{align}
  Z_{\rm{D0-D6}} &= e^{K/2}\,\left( p^0\,z^1\,z^2\,z^3 -q_0 \right) \equiv e^{K/2}\,Y_{\text{D0-D6}}~,
\end{align}
and
\begin{align}
  Z_1 =  e^{K/2}\,Y_1~, \quad Z_2 =  e^{K/2}\,\left(Y_{\text{D0-D6}}-Y_1\right)\,,
\end{align}
where we defined
\begin{align}
Y_1\equiv\left( P^0\,z^1\,z^2\,z^3-\frac{1}{2}s_{ABC}P^A\,z^B\,z^C + z^A\,Q_A -Q_0 \right)~.
\end{align}

Let us analyze the consequences due to the existence of a wall of
marginal stability on our general split described above.\footnote{At
this point, we assume that all charge vectors are supersymmetric. We
will study the requirements later.} From the condition
\eqref{eq:noim} :
\begin{equation}
  Z_1\,\bar{Z}_2 = \bar{Z}_1\,Z_2 \Leftrightarrow \frac{Z_1}{\bar{Z}_1} = \frac{Z_2}{\bar{Z}_2} \Leftrightarrow \frac{Y_1}{\bar{Y}_1}=\frac{Y_{\text{D0-D6}}-Y_1}{\bar{Y}_{\text{D0-D6}}-\bar{Y}_1} \Leftrightarrow \frac{Y_1}{\bar{Y}_1}= \frac{Y_{\text{D0-D6}}}{\bar{Y}_{\text{D0-D6}}}
 \label{eq:al1}
\end{equation}
Notice $\alpha_1= \alpha_2 + n\pi$, but also $\alpha_1 =
\alpha_{\text{D0-D6}} + m\pi$ for $n\,,m\in Z$. In other words, this
condition still allows both aligned and misaligned central charges.
Also the last equality would be perfectly consistent with an split of
the form $\Gamma_2 \rightarrow \Gamma_1 + \Gamma_{\text{D0-D6}}$,
which is not what we are interested in studying.

It is the second condition in \eqref{eq:noim}
\begin{equation*}
  \text{Re}\left(Z_1\,\bar{Z}_2\right) = |Z_1||Z_2|\,\cos(\alpha_1-\alpha_2) > 0\,,
\end{equation*}
that guarantees both split charges are aligned. Furthermore, since
$Z_2=Z_{\text{D0-D6}}-Z_1$, it follows
\begin{equation}
  |Z_2|\,e^{i\alpha_1} = \left((-1)^m\,|Z_{\text{D0-D6}}| - |Z_1|\right)\,e^{i\alpha_1}\,,
\end{equation}
which is only consistent when all three charges involved in the split
are aligned. Thus, it is the second condition in \eqref{eq:noim} that
breaks the reversibility of the split, disallowing channels such as
$\Gamma_2 \rightarrow \Gamma_1 + \Gamma_{\text{D0-D6}}$, since
\begin{equation*}
  M_{\rm D0-D6} = |Z_{\rm D0-D6}| =  M_1 + M_2 = |Z_1| + |Z_2|\,.
\end{equation*}

\subsection{U-duality orbits}

The D0-D6 system is a particular example of a state with negative
quartic invariant $(\Delta<0)$. Any other such state would be subject
to the same considerations discussed so far. Thus, it is important to
determine whether there exists any U-duality transformation relating
these different states so that the conclusions reached for the D0-D6
can be extended to the full subclass of these states. It was proved
in \cite{ferrara-latest} that all states with $\Delta < 0$ belong to
the same U-duality orbit. This is shown recalling that charges in the
STU model transform in the $(2,2,2)$ representation of the
$\left(\SL(2,\RR)\right)^3$ duality symmetry group. For completeness,
we include their proof below.

Let us parameterize the three $\SL(2,\RR)$ matrices building the
U-duality group $\left(\SL(2,\RR)\right)^3$ as
\begin{equation}
  M_A = \begin{pmatrix}
  a^A & b^A \\
  c^A & d^A \end{pmatrix}
  \quad {\rm with } \quad \det (M_A) = 1~, \quad A=1,2,3~.
\end{equation}
Consider a charge vector with arbitrary charges $(P^0,P^A;Q_A,Q_0)$
and the vector $(p^0,\,\vec{0};\,\vec{0},\,q_0)$. Given the
transformation properties of the charges, these two set of charges
are related by the set of constraints
\begin{eqnarray}
  -Q_0 &=& a^1\,a^2\,a^3\,q_0 + b^1\,b^2\,b^3\,p^0\,, \\
  Q_A &=& -\frac{1}{2}s_{BCD}\,c^B\,a^C\,a^D\,q_0 + \frac{1}{2}s_{BCD}\,d^B\,b^C\,b^D\,p^0\,, \\
  P^A &=& -\frac{1}{2}s_{BCD}\,a^B\,c^C\,c^D\,q_0 + \frac{1}{2}s_{BCD}\,b^B\,d^C\,d^D\,p^0\,, \\
  P^0 &=& c^1\,c^2\,c^3\,q_0 + d^1\,d^2\,d^3\,p^0\,.
\end{eqnarray}
This system is solved by the following set of matrices
\cite{ferrara-latest} :
\begin{equation*}
  M_A = -\frac{\text{sgn}(\xi)}{\sqrt{\left(\psi_A + \rho_A\right)\,\xi}}\,\begin{pmatrix}
  \psi_A\,\xi & - \rho_A \\
  \xi & 1 \end{pmatrix} \quad \Leftrightarrow \quad
  M_A^{-1} = - \frac{\text{sgn}(\xi)}{\sqrt{\left(\psi_A + \rho_A\right)\,\xi}}\,\begin{pmatrix}
  1 & \rho_A \\
  -\xi & \psi_A\,\xi \end{pmatrix} \,,
\end{equation*}
with
\begin{eqnarray}
  \xi &=& \left(\frac{p^0}{q_0}\right)^{1/3}\,\left(\frac{2P^1\,P^2\,P^3 + P^0\,\left(\sqrt{-\Delta}-P^\Lambda\,Q_\Lambda\right)}
  {2P^1\,P^2\,P^3 - P^0\,\left(\sqrt{-\Delta}-P^\Lambda\,Q_\Lambda\right)}\right)^{1/3} \in \RR\,, \\
  \psi_A &=& \frac{\sqrt{-\Delta} + P^\Lambda\,Q_\Lambda -2 P^A\,Q_A}{s_{ABC}\,P^B\,P^C - 2P^0\,Q_
  A} \in \RR\,\,(\text{no sum on A})\,, \\
  \rho_A &=& \frac{\sqrt{-\Delta} - P^\Lambda\,Q_\Lambda +2 P^A\,Q_A}{s_{ABC}\,P^B\,P^C - 2P^0\,Q_A} \in \RR\,\,(\text{no sum on A})\,.
\end{eqnarray}
These transformations preserve the value of $\Delta$, {\it i.e.}
\begin{equation*}
  \Delta = -(p^0\,q_0)^2 = -4Q_0\,P^1\,P^2\,P^3 -4P^0\,Q_1\,Q_2\,Q_3 -(P^\Lambda\,Q_\Lambda)^2 + 4\sum_{A<B} P^A\,Q_A\,P^B\,Q_B \,.
\end{equation*}
Thus, all the states in the orbit have negative quartic invariant.

As emphasized in \cite{ferrara-latest}, the above matrices are not
the most general ones that can be constructed connecting states with
negative quartic invariant. One could introduce a triple $\xi_A$
satisfying the constraint $\xi_1\,\xi_2\,\xi_3 = \xi^3$, a feature
that was already alluded to in the context of extremal non-BPS black holes in
\cite{GLS07}.

This result guarantees that given a wall of marginal stability and a pair of bound state constituents
in the D0-D6 frame, they also exist in any other frame related to the latter.

\section{Bound states of 1/4 and 1/2 BPS states}

In this section we will determine the pairs of 1/4 and 1/2 BPS
constituents that may form a bound state carrying only D0 and D6
brane charges by imposing local conditions on the system.\footnote{We
will discuss the possibility of 1/8 BPS constituents in a later
section. One could also consider n-state splits, but these are
necessarily co-dimension larger than one.} Our procedure is as
follows: first, we solve for all composite vectors $\Gamma_{1,2}$
that preserve at least 1/4 supercharges consistent with charge
conservation; second, we analyze whether the states associated with
such charge vectors exist; and finally, we derive the explicit
equations for the walls of marginal stability.


\subsection{Classification of final states}

Given a total charge vector $\Gamma_{{\rm
D0-D6}}=(p^0,\vec{0};\vec{0},q_0)$, we are looking for pairs of 1/4
and/or 1/2 BPS charge vectors $\{\Gamma_1,\,\Gamma_2\}$ such that
\begin{equation}\label{gda}
{\Gamma_{{\rm D0-D6}}\, = \,\Gamma_1+\Gamma_2}~,
\end{equation}
and with quartic invariant $\Delta$ satisfying \cite{Ferrara:1997ci}
\begin{equation}\label{gad}
{\Delta=0\quad {\rm and} \quad {\partial
\Delta\over\partial  q_\Lambda}=0~,\quad {\partial \Delta\over\partial
p^\Lambda}=0~. }
\end{equation}
In appendix \ref{sec:proof} we present a detailed derivation for the
general solution to these equations. There we argue that any charge
vector satisfying \eqref{gad} can be written as
\begin{align}\label{eq:genBPS}
\left(\beta_1P^0,\beta_2P^0,\alpha_1 P^2,\alpha_1 P^3;\beta_1Q_0,\alpha_2 P^3,\alpha_2 P^2,\beta_2Q_0\right)
\end{align}
with $\alpha_{1,2}$ and $\beta_{1,2}$ constants. Imposing the
conditions \eqref{gad} on \eqref{eq:genBPS} reduces to
\begin{equation}
\begin{array}{cc}\label{eq:cond}
P^0P^2P^3\,\alpha_{1,2}\,(\beta_1\alpha_2-\beta_2\alpha_1)&=0\,,  \\
Q_0P^2P^3\,\alpha_{1,2}\,(\beta_1\alpha_2-\beta_2\alpha_1)&=0\,, \\
Q_0P^0P^{2,3}\,\beta_{1,2}\,(\beta_1\alpha_2-\beta_2\alpha_1)&=0\,.
\end{array}
\end{equation}
Consider two such charge vectors $\{\Gamma_1,\,\Gamma_2\}$ consistent with charge conservation \eqref{gda} :
\begin{align}\label{gdb}
\Gamma_1&=(-ab p^0,-bcp^0, \alpha_1 p^2, \alpha_2 p^3; ad
q_0,\alpha_2 p^3, \alpha_2p^2, cd q_0)\\
\Gamma_2&=(cd p^0,bcp^0, -\alpha_1 p^2, -\alpha_2 p^3; -ad
q_0,-\alpha_2 p^3, -\alpha_2p^2, -ab q_0)
\end{align}
with
\begin{equation}\label{gdba}
{cd-ab=1~.}
\end{equation}
Both charge vectors $\Gamma_{1,2}$ must satisfy \eqref{eq:cond}.
Since $p^0,\,q_0\neq0$, conditions \eqref{eq:cond} on
$\Gamma_1$ reduce to
\begin{equation}\label{gde}
\begin{array}{cc}
p^2p^3\,(a\alpha_2-c\alpha_1)\,\alpha_{1,2}b&=0~,\\
p^2p^3\,(a\alpha_2-c\alpha_1)\,\alpha_{1,2}d&=0~,
\end{array}\quad\quad
\begin{array}{cc}
p^{2,3}\,(a\alpha_2-c\alpha_1)\,abd&=0~,\\
p^{2,3}\,(a\alpha_2-c\alpha_1)\,bcd&=0~,
\end{array}
\end{equation}
whereas for $\Gamma_{2}$ we have
\begin{equation}\label{gdea}
\begin{array}{cc}
p^2p^3\,(d\alpha_2-b\alpha_1)\,\alpha_{1,2}c&=0~,\\
p^2p^3\,(d\alpha_2-c\alpha_1)\,\alpha_{1,2}a&=0~,
\end{array}\quad\quad
\begin{array}{cc}
p^{2,3}\,(d\alpha_2-c\alpha_1)\,acd&=0~,\\
 p^{2,3}\,(d\alpha_2-c\alpha_1)\,abc&=0~.
\end{array}
\end{equation}
There are three ways to simultaneously solve \eqref{gde} and
\eqref{gdea}
\begin{align*}
i)&~~(a\alpha_2-c\alpha_1)=(d\alpha_2-b\alpha_1)=0~,\\
ii)& ~~(a\alpha_2-c\alpha_1)=0~~ \& ~~ (d\alpha_2-b\alpha_1)\neq0~; \quad (a\alpha_2-c\alpha_1)\neq0~~\&~~ (d\alpha_2-b\alpha_1)=0~,\\
iii)& ~~(a\alpha_2-c\alpha_1)\neq0~~\&~~ (d\alpha_2-b\alpha_1)\neq0~.
\end{align*}
For arbitrary values of $\{a,\,b,\,c,\,d\}$ satisfying \eqref{gdba}, condition $i)$
is only solved if $\alpha_1=\alpha_2=0$. These states, that we will refer to as
{\bf type I} states are
\begin{subequations}\label{final4}
\begin{align}
\Gamma^{(I)}_1&=(-ab p^0,[(-bcp^0)^A]; [(ad q_0)_A], cd q_0)\,,\label{final4a}\\
\Gamma^{(I)}_2&=(cd p^0, [(bc p^0)^A]; [(-adq_0)_A], -ab q_0)\,.\label{final4b}
\end{align}
\end{subequations}

For conditions $ii)$ and $iii)$, we find that the only non trivial solutions are obtained
by setting either $p^2$ (and/or $p^3$) and one of the
coefficients in \eqref{gdba} to zero. The resulting charge vectors, that we will refer to as
{\bf type II} states, are
\begin{subequations}\label{final8}
\begin{align}
\Gamma^{(II)}_1&=(p^0,[p^A]; [q_B], 0)\label{final8a}\\
\Gamma^{(II)}_2&=(0, [- p^A]; [-q_B], q_0)\label{final8b}
\end{align}
\end{subequations}
with $A\neq B$, and where the squared brackets are used
to denote that there is a single charge of the vector $p^A$ (or $q_A$)
turned on and the superscript (subscript) labels the component. For example,
$[p^1]=(p,0,0)$ and $[q_2]=(0,q,0)$.
%


\subsection{Existence of the split BPS states}\label{sec:existence}

Charge vectors \eqref{final4} and \eqref{final8} are supersymmetric,
but this does not guarantee the state carrying them exists. Since we
are eventually interested in interpreting two-centered supergravity
configurations as bound states composed of the states associated with
each center, we must first analyze when the individual states exist.
This is a difficult question, specially for states with $\Delta=0$,
but one requirement we implement is that their central charges do not
vanish. The conditions derived in this way match with the regularity
of $\Sigma^2$ in the single center supergravity realization of the
given state.

\paragraph{Type I states.} The central charges describing the charge vectors
\eqref{final4} are
\begin{subequations}\label{central4}
\begin{align}
Z^{(I)}_1&=-e^{K/2}\,(b\,p^0\,z^2z^3-d\,q_0)(a\,z^1-c)\label{central4a}~,\\
Z^{(I)}_2&=e^{K/2}\,(c\,p^0\,z^2z^3-a\,q_0)(d\,z^1-b)\label{central4b}~,
\end{align}
\end{subequations}
where we set $A=1$ in \eqref{final4}. Because of the factorized nature of these
central charges, their zeroes can occur in either of their factors.

$Z^{(I)}_1$ can vanish when $az_1= c$. This requires ${\rm Im}
z_1=0$, which is a singular point in moduli space, and lies beyond
the regime of validity of our supergravity approximation. The second
factor vanishes when $bp^0\,z^2\,z^3=dq_0$, which is a complex
equation. Assuming volumes never vanish, its imaginary part can be
solved by
\begin{equation*}
  B^2 = -\frac{J^2}{J^3}\,B^3\,.
\end{equation*}
Substituting this into its real part gives
\begin{equation*}
  -bp^0\,\frac{J^2}{J^3}\,|z^3|^2 = dq_0\,.
\end{equation*}
Notice that if any of the two parameters $\{b,\,d\}$ vanish, the
central charge will never vanish at a non-singular point in moduli
space. This corresponds to the particular cases of D6-D4 ($d=0$) and
D0-D2 ($b=0$). When $d,\,b\neq 0$, using the positivity of the
volumes and $p^0,\,q_0\neq 0$, we conclude:
\begin{itemize}
\item[1.] $Z^{(I)}_1$ has zeroes at non-singular points in moduli
    space {\it if} $bd\,p^0q_0 < 0$.
\item[2.] $Z^{(I)}_1$ has {\it no} zeroes at regular points
    in moduli space {\it if} $bd\,p^0q_0 > 0$.
\item[3.] $Z^{(I)}_1$ has {\it no} zeroes at regular points
    in moduli space whenever $d=0$ or $b=0$.
\end{itemize}

$Z^{(I)}_2$ has an analogous structure to the one for $Z^{(I)}_1$ and so are
the conclusions:
\begin{itemize}
\item[1.] $Z^{(I)}_2$ has zeroes at non-singular points in moduli
    space {\it if} $ac\,p^0q_0 < 0$.
\item[2.] $Z^{(I)}_2$ has {\it no} zeroes at regular points
    in moduli space {\it if} $ac\,p^0q_0 > 0$.
\item[3.] $Z^{(I)}_2$ has {\it no} zeroes at regular points
    in moduli space whenever $a=0$ or $c=0$.
\end{itemize}

From this analysis we conclude BPS constituents of type {\bf I} will
co-exist in the following cases:
\begin{itemize}
\item[1.] If $(a,\,b,\,c,\,d)$ are all non-vanishing, this requires
\begin{align}\label{cond4}
ac\,p^0q_0\,>0~, \quad bd\,p^0q_0\,>0~.
\end{align}
\item[2.] If $d=0$ ($ab=-1$) and $c\neq 0$, this requires
    $ac\,p^0q_0 > 0$ or equivalently $bc\,p^0q_0 <
    0$.\footnote{There is an analogous situation for $b=0$
    ($cd=1$) and $a\neq 0$, which also requires $ac\,p^0q_0 >
    0$.}
\item[3.] When $d=c=0$ ($ab=-1$) or $a=b=0$ ($cd=1$), the
    standard D0 + D6 split, constituents always exist.
\end{itemize}

It is interesting to relate these observations with the behavior of
the supergravity  solution near the charge source. All BPS states of
type {\bf I} have vanishing quartic invariant. The status of these
states as supergravity solutions to the attractor equations is less
obvious than those states having $\Delta > 0$ due to the singular
character of the solution at the pole (location of the charge).
Generically one needs to include higher order corrections in the
supergravity Lagrangian to properly describe these regions of
spacetime. Despite this fact, one should still demand a smooth
geometry at sufficiently large distance.

For example, BPS states carrying a single D-brane charge are
well-defined states that preserve half of the supercharges. In
particular their central charges never vanish on regular points of
the moduli space. In the supergravity approximation, this translates
into having $\Sigma^2$ positive throughout space-time even though the
size of the horizon is zero classically. For more general charge
vectors with $\Delta=0$, it is natural to analyze the behavior of the
factor $\Sigma^2$ as a function of the charges and moduli to
determine the existence of the state. The attractor mechanism only
fixes the value of $\Sigma^2$ at the horizon to be proportional to
$\Delta$. Thus, the dominant contribution to $\Sigma^2$ very close to
the charge source is no longer guaranteed to be independent of the
moduli, and the positivity of $\Sigma^2$ might not be satisfied.

Let us describe this more explicitly for generic type {\bf I} states.
Consider a single center BPS supergravity configuration realizing the
state with central charge $Z^{(I)}_1$ and charge vector
$\Gamma^{(I)}_1$. The phase of the central charge satisfies :
\begin{equation*}
  |Z^{(I)}_1|\,\sin\alpha = e^{K/2}\left(-abp^0{\rm Im}(z^1\,z^2\,z^3)
  +cbp^0{\rm Im}(z^2\,z^3)+adq_0\,J^1\right)\,.
\end{equation*}
For generic values of the parameters, the dominant contribution to
$\Sigma^2$ near the pole is given by
\begin{equation}
  \Sigma^2 (\vec{x}\to\vec{x}_1) \to
  16\,e^K\,J^2\,J^3\,\frac{|az^1-c|^2}{|\vec{x}-\vec{x}_1|^2}\,bdp^0q_0
  + {\cal O}\left(|\vec{x}-\vec{x}_1|^{-1}\right)
\label{eq:check1}
\end{equation}
Notice that \eqref{eq:check1} diverges as $1/r^2$, $r$ being the
distance to the pole, hence it corresponds to a 1/4 BPS state (see
\eqref{susy}). Furthermore, $\Sigma^2$ is {\it only} positive when
$bdp^0q_0 > 0$, which matches the condition derived from requiring
the absence of zeroes in $Z^{(I)}_1$ at regular points in moduli
space.

In the particular case $d=0$, the dominant contribution to $\Sigma^2$
in the same limit studied above is
\begin{equation}\label{eq:check2}
  \Sigma^2 (\vec{x}\to\vec{x}_1) \to \, |Z^{(I)}_1|\,\frac{4}{|\vec{x}-\vec{x}_1|} + {\cal O}(|\vec{x}-\vec{x}_1|^0)
\end{equation}
The $1/r$ divergence matches the 1/2 BPS character of this set of
states, and \eqref{eq:check2} is {\it always} positive in this limit,
in agreement with the regularity of the central charge for these
states.

What we would like to emphasize is that the precise value of the
dominant contribution to $\Sigma^2$ does depend on the moduli turned
on at infinity. In particular, it will generically depend on the
total central charge phase $\alpha$.

\paragraph{Type II split.} As for the type {\bf I} states, we want to determine if the
type {\bf II} states exist by demanding regularity of the central
charges associated to the states \eqref{final8}. For $A=1$ and $B=2$
in \eqref{final8}, the central charge for each state is
\begin{subequations} \label{central8}
\begin{align}
Z^{(II)}_1&=e^{K/2}\,(p^0z^2z^3-p^2z^3+q_1)\,z^1~,\label{central8a}\\
Z^{(II)}_2&=e^{K/2}\,(p^2z^1z^3-q_1z^1-q_0)~.\label{central8b}
\end{align}
\end{subequations}
According to \eqref{central8a} $Z^{(II)}_1$ has a factorized form,
its first factor $z_1$ only vanishing in singular points of moduli
space. If we focus on the second factor, its imaginary component
allows us to solve for one of the moduli:
\begin{equation*}
  p^0\,B^2 = p^2 - \frac{J^2}{J^3}\,B^3\,p^0\,.
\end{equation*}
Substituting this into the real part of the same factor, we obtain
the constraint
\begin{equation*}
  q_1 = p^0\,\frac{J^2}{J^3}\,|z^3|^2\,.
\end{equation*}
Since volumes $J^A$ are positive and $|z^3|$ only vanishes at
singular points in moduli space, we reach the conclusion:
\begin{itemize}
\item[1.] $Z^{(II)}_1$ has zeroes at non-singular points in
    moduli space {\it if} $p^0q_1 > 0$.
\item[2.] $Z^{(II)}_1$ has {\it no} zeroes at non-singular points
    in moduli space {\it if} $p^0q_1 < 0$.
\end{itemize}

The analysis for $Z^{(II)}_2$ is entirely analogous, and the
conclusions similar in nature:
\begin{itemize}
\item[1.] $Z^{(II)}_2$ has zeroes at non-singular points in
    moduli space {\it if} $p^2q_0 < 0$.
\item[2.] $Z^{(II)}_2$ has {\it no} zeroes at non-singular points
    in moduli space {\it if} $p^2q_0 > 0$.
\end{itemize}

Therefore, we have that a type {\bf II} state will be well defined in
moduli space if
\begin{equation}\label{cond8}
p^0q_1\,<\,0~, \quad p^2q_0\,>\,0~.
\end{equation}

Let us match these observations with the positivity of $\Sigma^2$
close to the pole, as we did for type {\bf I} states. Consider a
state with central charge $Z^{(II)}_1$ and charge vector
$\Gamma^{(II)}_1$. The central charge phase satisfies
\begin{equation*}
  |Z^{(II)}_1|\sin\alpha = {e^{K/2}}\,\left(p^0{\rm Im}(z^1\,z^2\,z^3)
   - p^2{\rm Im}(z^1\,z^3) + q_1\,J^1\right)\,.
\end{equation*}
The dominant contribution to $\Sigma^2$ close to the charge vector
location $(\vec{x}_1)$ is
\begin{equation}
  \Sigma^2 (\vec{x}\to\vec{x}_1) \to  \frac{16\,e^{K}\,J^2\,J^3\,|z^1|^2}{|\vec{x}-\vec{x}_1|^2}\,(-p^0q_1) + {\cal O}(|\vec{x}-\vec{x}_1|^{-1})
\end{equation}
The behavior is consistent with a 1/4 BPS state, as it should, and
the function is positive if $p^0q_1 < 0$, which matches the condition
for the regularity of $Z^{(II)}_1$.

\subsection{Walls of marginal stability}\label{sec:wall}

Having identified the potential BPS constituents for our bound
states, we would like to solve the conditions \eqref{eq:noim} for the
two possible splits: \eqref{final4} and \eqref{final8}. These
conditions describe walls of marginal stability and define the region
of moduli space where the bound states exist.

\paragraph{Type I split.} For this split, the D0-D6 charge vector decomposes into
\begin{align}
\Gamma_{\rm D0-D6} \quad \to \quad \Gamma_1^{(I)} +\Gamma_2^{(I)}~,
\end{align}
with the final states carrying charges \eqref{final4} and the central
charges of each constituent are \eqref{central4}. The imaginary and
real part of $\left(Z_1^{(I)}\bar{Z}_2^{(I)}\right)$ are given by
\begin{align}\label{imZ1Z24}
e^{-K}\,{\rm Im}\left({Z_1^{(I)}\bar{Z}_2^{(I)}}\right)= &-J^1\,\left(ad\,q_0^2-p^0q_0(ab+cd){\rm Re}(z^2z^3)+bc\,(p^0)^2|z^2z^3|^2\right)\nonumber\\
&-p^0q_0{\rm Im}(z^2z^3)\,\left(bc-(ab+cd)B^1+ad|z^1|^2\right)~,
\end{align}
and
\begin{align}\label{reZ1Z24}
e^{-K}\,{\rm Re}\left({Z_1^{(I)}\bar{Z}_2^{(I)}}\right)=
&-\left(bc-(ab+cd)B^1+ad|z^1|^2\right)\,
\left(ad\,q_0^2-p^0q_0(ab+cd){\rm Re}(z^2z^3)\right. \nonumber \\
& \left. +bc\,(p^0)^2|z^2z^3|^2\right) + p^0q_0{\rm Im}(z^2z^3)\,J^1~.
\end{align}
Imposing mass conservation,  $|Z_{\rm
D0-D6}|=|Z_{1}^{(I)}|+|Z_{2}^{(I)}|$, which is equivalent to setting
\eqref{imZ1Z24} equal to zero gives
\begin{align}\label{wall4}
 &J^1\,\left(ad\,q_0^2-p^0q_0(ab+cd){\rm Re}(z^2z^3)+bc\,(p^0)^2|z^2z^3|^2\right)\nonumber\\
 &=-p^0q_0{\rm Im}(z^2z^3)\,\left(bc-(ab+cd)B^1+ad|z^1|^2\right)~.
\end{align}
In addition, according to \eqref{eq:noim} the phases will be aligned
along the wall if ${\rm Re}{Z_1^{(I)}\bar{Z}_2^{(I)}}>0$, which
reduces to
\begin{equation}\label{phase4a}
p^0q_0{\rm Im}(z^2z^3)\,J^1>0~,
\end{equation}
where we used \eqref{reZ1Z24} and \eqref{wall4}. Since $J^1$ is
always positive and non-zero, \eqref{phase4a} becomes
\begin{equation}\label{phase4}
p^0q_0{\rm Im}(z^2z^3)\,>0~.
\end{equation}

Equation \eqref{wall4} describes circles or straight lines in the
$z^1$ complex plane for constant $(z^2z^3)$. These circles are
exactly those found in
\cite{Sen:2007vb,Cheng:2007ch,Sen:2007nz,Mukherjee:2007nc,Mukhi:2008ry},
where the analysis was done for 1/4 BPS states in ${\cal N}=4$ theory
decaying into two 1/2 BPS states. Here $z^1$ can be interpreted as
the axion-dilaton moduli. The charges vectors \eqref{final4} can be
written as electric ${Q}$ and magnetic ${P}$ vectors of the $O(6,n)$
duality group of ${\cal N}=4$. For example, if $A=1$ in
\eqref{final4a} the D-brane charges correspond in the Heterotic frame
to \cite{Dabholkar:2005dt}
\begin{equation}
\begin{split}\label{dyon}
Q^{(I)}&=(cdq_0,-bcp^0, \vec{0})~,\\
P^{(I)}&=(adq_0,-abp^0, \vec{0})~,
\end{split}
\end{equation}
and a similar expression for \eqref{final4b}. This is what we would
expect for 1/2 BPS states in ${\cal N}=4$, since the electric and
magnetic vectors in \eqref{dyon} are parallel.

\paragraph{Type II split.} We proceed to determine the marginal stability condition for D0-D6 when final
states carry the charges in \eqref{final8}. For simplicity, we
re-write the central charge of each constituent \eqref{central8} as
\begin{subequations} \label{modII}
\begin{align}
Z^{(II)}_1&=e^{K/2}\,(p^0z^1z^2z^3-Y)~,\label{modIIa}\\
Z^{(II)}_2&=e^{K/2}\,(Y-q_0)~.\label{modIIb}
\end{align}
\end{subequations}
with
\begin{align}
Y\equiv p^2z^1z^3-q_1z^1~.
\end{align}
The imaginary and real part of
$\left(Z_1^{(II)}\bar{Z}_2^{(II)}\right)$ are
\begin{align}\label{imZ1Z28}
e^{-K}\,{\rm Im}\left({Z_1^{(I)}\bar{Z}_2^{(I)}}\right)= &{\rm Im}Y\left(q_0-p^0{\rm Re}(z^1z^2z^3)\right)
+p^0{\rm Im}(z^1z^2z^3)\left({\rm Re}Y-q_0\right)~,
\end{align}
and
\begin{align}\label{reZ1Z28}
e^{-K}\,{\rm Re}\left({Z_1^{(I)}\bar{Z}_2^{(I)}}\right)=
-\left(q_0-p^0{\rm Re}(z^1z^2z^3)\right)\left({\rm Re}Y-q_0\right)
+p^0{\rm Im}(z^1z^2z^3){\rm Im}Y-|Y-q_0|^2
&~.
\end{align}
The first condition of marginal stability in \eqref{eq:noim}
simplifies to
\begin{align}\label{wallII}{\rm Im}Y\left(q_0-p^0{\rm Re}(z^1z^2z^3)\right)
+p^0{\rm Im}(z^1z^2z^3)\left({\rm Re}Y-q_0\right)=0~,
\end{align}
which imposes mass conservation. The phase of each state will be
align along the wall \eqref{wallII} when
\begin{align}\label{phaseII}
|Y-q_0|^2\left(-1+{p^0{\rm Im}(z^1z^2z^3)\over{\rm Im}Y}\right)\,>\,0~.
\end{align}
The two conditions, \eqref{wallII} and \eqref{phaseII}, define the
wall of marginal stability for type {\bf II} bound states. In the
following section, we will investigate if the conditions found in
this section for type {\bf I} and {\bf II} splits are sufficient or
just necessary for the state to have a well-behaved supergravity
description.

%
%
%
%
%
%
%


\section{Bound states as two-centered solutions}\label{sec:regsugra}

We will now examine whether the actual bound state, when realized as
a two-centered supergravity configuration, is a regular
configuration. Previously, we established a set of possible charge
splits of the total D0-D6 charge vector consistent with
supersymmetry, and  we described the regions of moduli space where
the individual and bound BPS states exist by imposing local algebraic
conditions. In the following we will study global conditions on the
geometry to assure the existence of the bound state.

\subsection{D0 and D6 constituents}

To illustrate the procedure we start with the simplest bound state,
{\it i.e.} $\Gamma_{\rm D0-D6}=\Gamma_{\rm D0}+\Gamma_{\rm D6}$. This
corresponds to $c=d=0$ and $ab=-1$ in \eqref{final4}. The constituent
central charges are
\begin{align}
Z_{\rm D6}=e^{K/2}\,p^0z^1z^2z^3~,\quad Z_{\rm D0}=-e^{K/2}\,q_0~.
\end{align}
Both quantities are regular in non-singular points of moduli space.
The metric and one form $\omega$ are as discussed in section
\ref{sec:BPSsol}, and the helicity of the state is $\langle
\Gamma_{\rm D6},\Gamma_{\rm D0}\rangle = -p^0q_0$. We choose the
D6-branes to be located at the origin $\vec{x}_1=\vec{0}$ with charge
$p^0$, and the D0 branes at $\vec{x}_2=(0,0,R)$ with charge $q_0$.
The set of harmonic functions are
\begin{align}\label{Hd0d6}
H^0=h^0+{p^0\over r}~, \quad H_0=h_0+{q_0\over \Theta}~, \quad H_A=h_A~, \quad H^A=h^A~,
\end{align}
with $r$ and $\Theta$ defined by \eqref{polar}. The integrability
conditions \eqref{int} reduce to
\begin{align}\label{intd0d6}
p^0h_0=q_0h^0~,\quad {p^0q_0\over R}=-h^0q_0~.
\end{align}
Using the  moduli identities listed in appendix \ref{sec:moduli} and
the integrability conditions \eqref{intd0d6}, the function
\eqref{sigma} reads
\begin{align}\label{warp}
\Sigma^2(H)=&-{1\over r^2}\left({p^0q_0\over R}\right)^2\left[1+{r-R\over \Theta}\right]^2
+{4\over r\Theta}\left({p^0q_0}\right)(h^1h_1+4e^{K}B^1J^2J^3)\nonumber\\
&+{4\over |Z_{\rm D0D6}|}\left[{1\over \Theta}{\rm Re}(Z_{\rm D0-D6}\bar{Z}_{D6})
+{1\over r}{\rm Re}(Z_{\rm D0-D6}\bar{Z}_{\rm D0})\right]+1\,.
\end{align}
The existence of the bound state requires that \eqref{warp} is
positive definite throughout spacetime. In particular, close to each
center we have
\begin{align}\label{poleD0D6}
\Sigma^2(\vec{x}\to\vec{x}_1)&=-{4p^0h_1\over  r }\left(-{q_0\over R}h^1+4e^{K}|z^1|^2J^2J^3\right) + \ldots~,\\
\Sigma^2(\vec{x}\to\vec{x}_2)&=-{4q_0h^1\over  \Theta }\left(-{p^0\over R}h_1+4e^{K}J^2J^3\right) + \ldots \nonumber ~,
\end{align}
where the dots denote subleading terms. Notice the divergence at each
center is consistent with having a 1/2 BPS charge vector constituent,
but the actual coefficient does depend on the moduli and the {\it
total} central charge phase $\alpha$. Contrary to what occurs for
single centered 1/2 BPS supergravity configurations in
\eqref{eq:check1} and \eqref{eq:check2}, the above expressions are
not positive definite for {\it any} value of the moduli and $\alpha$.
The analysis of marginal stability in section \ref{sec:wall}, showed
that the phases of the central charges are aligned if $p^0q_0{\rm
Im}(z^2z^3)>0$. Combining this with $J^{A}>0$ and \eqref{app:modh1}
tells us that
\begin{align}
-h^1q_0\,>\,0~\quad -p^0h_1\,>\,0~.
\end{align}
Therefore, the near pole behavior \eqref{poleD0D6} is positive in the
same region of moduli space described by the conditions of existence of the bound state in the
previous section.

Further, one can prove the absence of CTCs in the full geometry by
proving that \eqref{eq:ctc} is satisfied everywhere. For the D0-D6
bound state we have
\begin{align}\label{ctcd0d6}
\Sigma^2r^2\sin\theta^2-\left(\omega_\phi\right)^2=
&\sin^2\theta\Bigg[{4r\over \Theta}p^0q_0\,e^{K}\left({\rm Im}(z^2z^3)J^1
+B^1 J^2J^3\right)\nonumber\\
&+{4r^2\over |Z_{\rm D0D6}|}\left({1\over \Theta}{\rm Re}(Z_{\rm D0-D6}\bar{Z}_{D6})
+{1\over r}{\rm Re}(Z_{\rm D0-D6}\bar{Z}_{\rm D0})\right) +r^2\Bigg]
\\&+{4\over \Theta}\left({p^0q_0\over R}\right)^2(1-\cos\theta)(r+R-\Theta)~.\nonumber
\end{align}
Each term in  \eqref{ctcd0d6} is positive definite in the region of
moduli space defined by $\langle \G_1,\G_2\rangle {\rm
Im}(Z_{1}\bar{Z}_{2})>0$ and ${\rm
Re}(Z_{1}\bar{Z}_{2})>0$.\footnote{The first term proportional to
$p^0q_0\left({\rm Im}(z^2z^3)J^1+B^1 J^2J^3\right)$ can be shown to
be positive by assuming it is negative and then showing such an
assumption is not consistent with $\langle \G_1,\G_2\rangle {\rm
Im}(Z_{1}\bar{Z}_{2})>0$ and ${\rm Re}(Z_{1}\bar{Z}_{2})>0$.} Thus,
the conditions of marginal stability are sufficient for a regular
two-centered solution to exist with D0 and D6 charge split.

In the remaining of this section we will study the regularity of the
supergravity configurations describing the more general type {\bf I}
and type {\bf II} split states identified before. The tools and
methodology are the same as for the D0-D6. We will argue that for
only very specific cases the conditions of marginal stability
\eqref{eq:noim} are sufficient to guarantee regularity of the
two-centered solution.

\subsection{Type I bound states}

Consider a two-centered configuration with centers
$\vec{x}_1=\vec{0}$ and $\vec{x}_2=(0,0,R)$ carrying charges
$\Gamma_1^{(I)}$ and $\Gamma_2^{(I)}$, respectively. For simplicity,
we will set $A=1$ in \eqref{final4}. The set of harmonic functions is
given by
\begin{align*}
H^0=h^0-{ab\,p^0\over r} +{cd\,p^0\over \Theta}~, \quad & \quad
H^1=h^1-{bc\,p^0\over r} +{bc\,p^0\over \Theta}~,\\
H_0=h_0+{cd\,q_0\over r} -{ab\,q_0\over \Theta}~, \quad & \quad
H_1=h_1+{ad\,q_0\over r} -{ad\,q_0\over \Theta}~,
\end{align*}
with $cd-ab=1$ and $\Theta^2=r^2+R^2-2rR\cos\theta$. The remaining
harmonic functions are constant, {\it i.e.} $H^{2,3}=h^{2,3}$ and
$H_{2,3}=h_{2,3}$. The factor \eqref{sigma} is
\begin{align}\label{sigma4}
\Sigma^2(H) = &-\left(-H_0H^0+H_1H^1+h_2h^2+h_3h^3\right)^2+4H^1H_1\left(h^2h_2+h^3h_3\right)\nonumber\\
 &-4H^0H_1h_2h_3-4H_0H^1h^2h^3 +4h^2h_2h^3h_3 ~.
\end{align}
The integrability conditions \eqref{int} read
\begin{equation}\label{mod4}
q_0h^0=p^0h_0~,\quad {p^0q_0\over R}=-q_0h^{0}+{1\over ab+cd}\left(ad\,q_0h^1+bc\,p^0h_1\right)~,
\end{equation}
whereas the helicity of the state is
given by
\begin{equation}
\langle \Gamma_1^{(I)}, \Gamma_{2}^{(I)}\rangle= (ab+cd)p^0q_0~.
\end{equation}

Now we proceed to study the positivity of \eqref{sigma4}. As
$\vec{x}\to \infty$ the metric is asymptotically flat, therefore
$\Sigma^2\to 1$. Close to each pole $\{\vec{x}_1, \vec{x}_2\}$ it
should remain positive in order to avoid fake horizons. In the limit
$\vec{x}\to \vec{x}_1$, the leading terms in \eqref{sigma4} are
\begin{align}\label{4close}
\Sigma^2(\vec{x}\to \vec{x}_1) = &-{1\over r^2}
\left(-q_0h^0-{p^0q_0\over R}+ad\,q_0h^1-bc\,p^0h_1\right)^2
-{4abcd\,p^0q_0\over r^2}\left(h^2h_2+h^3h_3\right)\nonumber\\
 &+{4a^2bd\,p^0q_0\over r^2}h_2h_3+{4b^2cd\,p^0q_0\over r^2}h^2h^3
 +{\cal O}\left({1\over r}\right) ~.
\end{align}
Using \eqref{mod4} and after some algebra, we can rewrite
\eqref{4close} as
\begin{align}\label{4aclose}
\Sigma^2(\vec{x}\to \vec{x}_1) = &-{4bdp^0q_0\over r^2}
\left(a^2\,q_0h^1+c^2\,p^0h_1\right){1\over R}
+{4abcd\over r^2}\left({p^0q_0\over R}+h^0q_0\right)^2\nonumber\\
 &+{4bd\,p^0q_0\over r^2}\left[a^2(h_2h_3-h_0h^1)+ c^2(h^2h^3-h^0h_1)-ac\left(h^2h_2+h^3h_3-h_1h^1\right)\right]
 \nonumber\\&+{\cal O}\left({1\over r}\right) ~.
\end{align}
Notice the dependence on the moduli and the total central charge
phase $\alpha$ is very different from the one we found for the single
centered solution with the same center vector charge in
\eqref{eq:check1}. This is because of the singular nature of these
solutions to the attractor equations. Since $\Delta=0$, the dominant
(non-vanishing) contribution to $\Sigma^2$ is not fixed by the
attractor mechanism, and as such, it depends on global aspects of the
solution. From this perspective, the positivity of $\Sigma^2$ at each
center is already a non-trivial condition for the bound state to
exist.

Analogously, the behavior of \eqref{sigma4} close to the second
center is
\begin{align}\label{4bclose}
\Sigma^2(\vec{x}\to \vec{x}_2) = &-{4acp^0q_0\over \Theta^2}
\left(b^2\,q_0h^1+d^2\,p^0h_1\right){1\over R}
+{4abcd\over \Theta^2}\left({p^0q_0\over r_{12}}+h^0q_0\right)^2\nonumber\\
 &+{4ac\,p^0q_0\over \Theta^2}\left[d^2(h_2h_3-h_0h^1)+ b^2(h^2h^3-h^0h_1)-bd\left(h^2h_2+h^3h_3-h^1h_1\right)\right]
 \nonumber\\&+{\cal O}\left({1\over \Theta}\right) ~.
\end{align}

From the condition of marginal stability, we found that the bound
state will exist when \eqref{phase4} holds. Combining this condition
with the fact that $J^A>0$ in \eqref{app:modh1}, we have
\begin{align}
-h^1q_0\,>\,0~\quad -p^0h_1\,>\,0~.
\end{align}
Therefore the first and second term in \eqref{4aclose} and
\eqref{4bclose} will be positive if $bdp^0q_0>0$ and $acp^0q_0>0$.
This is consistent with the condition \eqref{cond4} derived by imposing
regularity of the central charge vectors.

Using \eqref{mod4}, \eqref{app:mod4a} and \eqref{app:modhh}, we can
write the last term in \eqref{4aclose} and \eqref{4bclose} as
\begin{subequations}\label{poleI}
\begin{align}
\Sigma^2(\vec{x}\to \vec{x}_1) = &\ldots+{4bd\,p^0q_0\over r^2}\,e^KJ^2J^3|az_1-c|^2
-{4abcd\over r^2}\left(h^0q_0\right)^2
+ {\cal O}\left({1\over r}\right)~,\\
\Sigma^2(\vec{x}\to \vec{x}_2) = &\ldots+{4ac\,p^0q_0\over \Theta^2}\,e^KJ^2J^3|dz_1-b|^2
-{4abcd\over \Theta^2}\left(h^0q_0\right)^2
+ {\cal O}\left({1\over \Theta}\right)~.
\end{align}
\end{subequations}
The first term for both poles is also positive if  $bdp^0q_0>0$ and
$acp^0q_0>0$, but the second term is negative for this assignment of
charges. This tell us that in order to have $\Sigma^2>0$ for
$(a,b,c,d)$ non-zero we need to impose further constraints on the
moduli, which will raise the co-dimension of the walls of marginal
stability. This conclusion can be avoided if we have one (or two) vanishing coefficients
among $(a,b,c,d)$ while still satisfying $cd-ab=1$. In these cases, the bound state may still
exist. We will explore in more detail this scenario in the remaining of this section.

Before proceeding, let us emphasize that at this point we have already established the existence of further requirements beyond supersymmetry, regularity of the central charge and existence of a wall of marginal stability for the supergravity supersymmetric bound state to exist. From a purely supergravity perspective, this also provides an example for families of configurations that solve the integrability conditions but are {\it not} free of CTCs.

\subsubsection{Surviving Type I states}

For non-zero values of $(a,b,c,d)$, we found in \eqref{poleI} that
the conditions of marginal stability are not sufficient to assure a
positive $\Sigma^2$ close to each pole. But if one of the integers is
zero, the negative contribution in \eqref{poleI} vanishes. In the
following, we will study the regularity of the supergravity solutions
for such configurations. Consider
\begin{align}\label{subtypeIa}
\Gamma_1&=(p^0,[-p];0,0)~,\quad \Gamma_2=(0,[p];0,q_0)~,
\end{align}
where the first vector corresponds to a D6 brane ($p^0$) and an
anti-D4 wrapping a 4-cycle of $T^6$ with charge $-p$, and the vector
$\Gamma_2$ corresponds to a D0 brane ($q_0$) and a D4 wrapping the
same cycle. The other possible combination is
\begin{align}\label{subtypeIb}
\Gamma_1&=(p^0,0;[q],0)~,\quad \Gamma_2=(0,0;[-q],q_0)~,
\end{align}
where the first vector corresponds to a D6 brane  and a D2 wrapping a
2-cycle of $T^6$ with charge $q$, and the vector $\Gamma_2$
corresponds to a D0 brane and an anti-D2 wrapping the same cycle.
Using the notation in \eqref{final4}, states \eqref{subtypeIa}
correspond to $d=0$ and $p\equiv bcp^0$, and states \eqref{subtypeIb}
correspond to $c=0$ and $q\equiv adq^0$.

The analysis of regularity for both configurations \eqref{subtypeIa}
and \eqref{subtypeIb} is completely analogous. For brevity, we will
carry the analysis only for \eqref{subtypeIa}. First consider
the conditions of marginal stability. The central charges of each
state is given by
\begin{align}\label{centralD4}
Z_{1}=e^{K/2}\left(p^0\,z^1+p\right)\,z^2\,z^3~,\quad Z_{ 2}=-e^{K/2}\left(p\,z^2\,z^3+q_0\right)~.
\end{align}
Demanding regularity of the central charges requires
\begin{align}\label{signD4}
p\,q_0\,<\,0~.
\end{align}
The bound state is stable if
\begin{align}\label{eq:stableD4}
  \langle \Gamma_1,\Gamma_2\rangle{\rm Im}\left(Z_1\,\bar{Z}_2\right) > 0\,,\quad
  {\rm Re}\left(Z_1\,\bar{Z}_2\right) > 0\,.
\end{align}
Inserting \eqref{centralD4} in the above conditions\footnote{Or
equivalently setting $d=0$ in \eqref{wall4} and \eqref{phase4}.}
gives
\begin{align}\label{boundD4}
{\rm Im}z^1{\rm Re}(z^2z^3)+ {\rm Im}(z^2z^3) {\rm Re}z^1+{p\over p^0}{\rm Im}(z^2z^3)+{p\over q_0}|x|^2{\rm Im}z^1>0~,
\end{align}
and
\begin{align}\label{boundD4a}
p^0q_0{\rm Im}(z^2z^3)>0~.
\end{align}

We proceed now to investigate the regularity conditions of the
supergravity solution. One important requirement is the absence of
closed timelike curves
\begin{align}\label{ctcD4}
\Sigma^2r^2\sin\theta^2-\left(\omega_\phi\right)^2\,>\,0~.
\end{align}
If \eqref{ctcD4} is satisfied this will also imply that $\Sigma^2$ is
positive through out the geometry. For the solution in hand, the
metric factor \eqref{sigma4} is
\begin{align}\label{sigmaD4}
\Sigma^2(H)=&~1+{4\over |Z_{\rm D0D6}|}\left({1\over r}{\rm Re}(Z_1\bar{Z}_{\rm D0-D6})+{1\over \Theta}{\rm Re}(Z_2\bar{Z}_{\rm D0-D6})\right)
\nonumber\\
&+{4\over r\Theta}{p^0q_0}\,e^{K}\left({\rm Im}(z^2z^3)J^1+B^1J^2J^3+{p\over p^0}J^2J^3\right)
+{4\over r\Theta^2}{(pq_0)}p^0h_1\nonumber\\
&+{4\over r\Theta}\left({p^0q_0\over R}\right)^2 -{1\over r^2}\left({p^0q_0\over R}\right)^2\left(1+{r-R\over\Theta}\right)^2
\end{align}
where $Z_{1,2}$ are defined by \eqref{centralD4}. The one-form
rotation is given by \eqref{omega} and for the charges
\eqref{subtypeIa} it reads
\begin{align}\label{omegaI}
\omega=-{p^0q_0\over R}
\left[1-{r+R\over \Theta}\right](1-\cos\theta)d\phi~.
\end{align}
Inserting \eqref{sigmaD4} and \eqref{omegaI} in \eqref{ctcD4} we get
\begin{align*}
\Sigma^2r^2\sin^2\theta-\left(\omega_\phi\right)^2=& r^2\sin^2\theta\Big[
1+{4\over |Z_{\rm D0D6}|}\left({1\over r}{\rm Re}(Z_1\bar{Z}_{\rm D0-D6})+{1\over \Theta}{\rm Re}(Z_2\bar{Z}_{\rm D0-D6})\right)\nonumber\\
&+{4\over r\Theta}{p^0q_0}\,e^K\left({\rm Im}(z^2z^3)J^1+B^1J^2J^3+{p\over p^0}J^2J^3\right)
+{4\over r\Theta^2}{(pq_0)}p^0h_1
\Big]
\\&+{4\over \Theta}\left({p^0q_0\over R}\right)^2(1-\cos\theta)(r+R-\Theta)
\end{align*}

All terms are positive definite. Thus, these configurations are free
of CTCs. As a consequence of this derivation, $\Sigma^2$ is positive
everywhere, and we conclude the supergravity realization of the
supersymmetric bound state exists.

\subsection{Type II split}

The discussion is analogous to type I. The bound state should be a
two-centered solution with centers $\vec{x}_1$ and $\vec{x}_2$
carrying charges $\Gamma_1^{(II)}$ and $\Gamma_2^{(II)}$ given by
\eqref{final8}. The set of harmonic functions are given by
\begin{align*}
H^0&=h^0+{p^0\over |\vec{x}-\vec{x}_1|}~, &H^2=h^2+{p^2\over |\vec{x}-\vec{x}_1|} -{p^2\over |\vec{x}-\vec{x}_2|}~,\\
H_0&=h_0+{q_0\over |\vec{x}-\vec{x}_2|} ~,
&H_1=h_1+{q_1\over |\vec{x}-\vec{x}_1|} -{q_1\over |\vec{x}-\vec{x}_2|}~.
\end{align*}
The remaining harmonic functions are constant, {\it i.e.}
$H^{1,3}=h^{1,3}$ and $H_{2,3}=h_{2,3}$. From \eqref{sigma}, the
metric factor for this bound state is
\begin{align}\label{sigma8}
\Sigma^2(H) = &-\left(-H_0H^0+H_1h^1+h_2H^2+h_3h^3\right)^2+4h^1H_1\left(H^2h_2+h^3h_3\right)\nonumber\\
 &-4H^0H_1h_2h_3-4H_0h^1H^2h^3 +4H^2h_2h^3h_3 ~.
\end{align}
For the charge vectors \eqref{final8}, the helicity of the state is
\begin{equation}
\langle \Gamma_1^{(II)}, \Gamma_{2}^{(II)}\rangle= -p^0q_0~,
\end{equation}
and the integrability conditions \eqref{int} reduce to
\begin{equation}\label{mod8}
q_0h^0=p^0h_0~,\quad {p^0q_0\over R}=-q_0h^{0}-q_1h^1+p^2h_2~.
\end{equation}

As before, our first check is to study the positivity of $\Sigma^2$
close to each pole. In the limit $\vec{x}\to \vec{x}_1$ and
$\vec{x}\to \vec{x}_2$, the leading terms in \eqref{sigma8} are
\begin{subequations}\label{8close}
\begin{align}
\Sigma^2(\vec{x}\to \vec{x}_1) &= -{4q_1p^0\over |\vec{x}-\vec{x}_1|^2}
\left(4e^K|z^1|^2J^2J^3-{q_0h^1\over R}\right)
 +{\cal O}\left( |\vec{x}-\vec{x}_1|^{-1}\right) ~,\label{8closea}\\
 \Sigma^2(\vec{x}\to \vec{x}_2) &= {4q_0p^2\over |\vec{x}-\vec{x}_2|^2}
\left(4e^KJ^2J^3-{p^0h_2\over R}\right)
 +{\cal O}\left(|\vec{x}-\vec{x}_2|^{-1}\right) ~.\label{8closeb}
\end{align}
\end{subequations}
where we used \eqref{mod8} and \eqref{app:mod4a}. From section
\ref{sec:existence}, the central charges $Z_1^{(II)}$ and $Z_2^{(II)}$ are regular if
\begin{align}
p^0q_1\,<\,0~,\quad  p^2q_0\,>\,0~,
\end{align}
hence the first term in each parenthesis in \eqref{8close} is positive.
The second term in \eqref{8closea} gives
\begin{align}\label{q1h1}
q_0p^0q_1h^1={2e^{K/2}\over |Z_{\rm D0-D6}|}\left(-p^0q_1(q_0)^2J^1-p^0q_1(q_0p^0)|z^1|^2{\rm Im}(z^2z^3)\right)~.
\end{align}
Is this quantity positive? From the analysis of the central charges
and the integrability conditions, the stable region for the state is
defined by $R>0$ in \eqref{mod8} and delimited by the walls of
marginal stability \eqref{wallII} and \eqref{phaseII}. These
conditions are not sufficient for having \eqref{q1h1} positive
definite. Analogously, by studying \eqref{8closeb} we reach the same
result. Therefore we conclude that $\Sigma^2$ can be negative close
to the poles unless we impose further constraints on the moduli,
increasing the co-dimension of the walls of marginal stability.

\section{Bound states including 1/8 BPS states}

In previous sections, we studied the supersymmetric D0-D6 bound
states as supergravity two-centered configurations involving 1/4 and
1/2 BPS charge vectors. In principle, it is also possible to include
as constituents 1/8 BPS states with vanishing quartic
invariant.\footnote{The possibility of allowing 1/8 BPS states with
positive quartic invariant is entropically disfavored, and we will
not consider it here.} Here we will argue that regularity of the
solution will generically impose further constraints on the moduli.
Thus, if it exists, it will do so in a region of moduli space of
co-dimension higher than one. Our strategy consists on studying the
behavior of $\Sigma$ for a generic two-centered solution, where one
of the centers is a 1/8 BPS state with vanishing quartic invariant.
Close to this center the positivity of $\Sigma^2$ is not guaranteed
by the integrability conditions, hence generically there will be
additional restrictions on the moduli.

Consider a two-centered supergravity configuration such that $
\Gamma_{\text{D0-D6}}=\Gamma_1+ \Gamma_2$. The pole at $\vec{x}_1$
carries a charge vector $\Gamma_1=(p^\Lambda_1,q_\Lambda^1)$
corresponding to a 1/8 BPS state with vanishing quartic invariant.
Thus, $\Delta_1=0$ and at least one $\partial\Delta_1/\partial
p^\Lambda$ and/or $\partial\Delta_1/\partial q_A$ are non-vanishing.
The second pole $\vec{x}_2$ carries charge
$\Gamma_2=(p^\Lambda_2,q_\Lambda^2)$.\footnote{Since $\Gamma_2$ is
supersymmetric and has vanishing quartic invariant, $\Delta_2=0$,
conservation of charge puts some non-trivial constraints on its
components. We will not need these details here, though this is a
problem that can be solved.} The behavior of the function $\Sigma^2$
close to the center $\vec{x}_1$ is
\begin{multline}
  \Sigma^2(\vec{x}\to\vec{x}_1) \to \frac{1}{|\vec{x}-\vec{x}_1|^3}\,
  \left(\frac{\partial\Delta_1}{\partial p^0_1}\,\left(h^0 + {p^0_2\over R}\right) +
  \frac{\partial\Delta_1}{\partial p^A_1}\,\left(h^A + {p^A_2\over R}\right) \right. \\
  \left. +  \frac{\partial\Delta_1}{\partial q_0^1}\,\left(h_0 + {q^2_0\over R}\right)
  + \frac{\partial\Delta_1}{\partial q_A^1}\,\left(h_A + {q^2_A\over R}\right)\right) +
  {\cal O}\left(|\vec{x}-\vec{x}_1|^{-2}\right)~.
\end{multline}
Given its linear dependence on $(h^\Lambda;\,h_\Lambda)$, we can use
the integrability condition \eqref{int} fixing the distance scale
$R$ between the two centers and the definitions given in
\eqref{moduli} to rewrite this expression as
\begin{equation}
   \Sigma^2(\vec{x}\to\vec{x}_1) \to  \frac{2\,{\rm Im}\left(Z_{\star}\,\bar{Z}_{\text{D0-D6}}\right)}{|Z_{\text{D0-D6}}|\,\langle\G_1,\,\G_2\rangle}
   \,{1\over |\vec{x}-\vec{x}_1|^3} + {\cal O}\left(|\vec{x}-\vec{x}_1|^{-2}\right)~,
\end{equation}
where $Z_{\star}$ is the central charge associated with the effective
charge vector
\begin{align}\label{Gstar}
  \Gamma_\star &= \langle\Gamma_1,\,\Gamma_2\rangle\, \Gamma_{\text{eff}} - \langle\Gamma_{\text{eff}},\,\Gamma_2\rangle\,\Gamma_1\,,
\end{align}
with
\begin{align*}
  \Gamma_{\text{eff}}& \equiv \left(\frac{\partial\Delta_1}{\partial q_0^1},\,-\frac{\partial\Delta_1}{\partial q_A^1}
  ;\,\frac{\partial\Delta_1}{\partial p^A_1},\,-\frac{\partial \Delta_1}{\partial p^0_1}\right)\,.
\end{align*}
Thus, positivity of $\Sigma^2$ in this limit requires
\begin{equation}
  \langle\G_1,\,\G_2\rangle\,{\rm Im}\left(Z_{\star}\,\bar{Z}_{\text{D0-D6}}\right) > 0\,.
 \label{posreq}
\end{equation}
Generically, this imposes a condition on the relative phases of both
central charges, which is moduli dependent.

Let us assume the existence of a supersymmetric bound state in a
region of moduli space bounded by a wall of marginal stability of
co-dimension one. This requires the following conditions to hold
\begin{equation}
  \langle \Gamma_1,\,\Gamma_{\text{D0-D6}} \rangle\,{\rm Im}\left(Z_1\,\bar{Z}_{\text{D0-D6}}\right) > 0 \quad \text{and} \quad {\rm Re}\left(Z_1\,\bar{Z}_2\right) > 0\,.
 \label{eq:18bound}
\end{equation}
The question is whether \eqref{eq:18bound} guarantees the positivity
of $\Sigma^2$ at the center $\vec{x}_1$ without introducing any
further constraint on the moduli, {\it i.e.} if \eqref{posreq} is
consistent with \eqref{eq:18bound}. A subset of effective central
charges $Z_{\star}$ that would trivially satisfy this property would
be
\begin{align*}
 Z_{\star} = \left(\beta +
i\,\langle\G_1,\,\G_2\rangle\,\gamma\right)\,Z_1 + \left(\alpha +
i\,\langle\G_1,\,\G_2\rangle\,\delta\right)\,Z_{2}~,
\end{align*}
$\forall\, \alpha<0$ and $\forall\,\beta,\,\gamma,\,\delta> 0$. This
imposes a condition on the effective charge vector
$\Gamma_{\star}$,\footnote{$\Gamma_\star$ does not have to correspond
to any physical charge in principle. It is just a convenient
mathematical way of encoding the behavior of $\Sigma^2$ near the pole
$\vec{x}_1$.}
\begin{equation}
  \Gamma_\star =  \left(\beta + i\,\langle\G_1,\,\G_2\rangle\,\gamma\right)\,\G_1 + \left(\alpha + i\,\langle\G_1,\,\G_2\rangle\,\delta\right)\,\G_{2}\,,
 \label{eq:eff}
\end{equation}
which is a non-linear equation to be satisfied for the charge
components of the original 1/8 BPS state. Since this is an equality
between charge vectors, we can check its consistency with charge
conservation by computing its inner product with $\Gamma_1$ and
$\Gamma_2$. Using the fact that $\langle\G_{\text{eff}},\,\G_1\rangle
= -4\Delta_1 = 0$, we learn from \eqref{Gstar} that
\begin{align*}
\langle\G_1,\,\G_\star\rangle = 0~,
\end{align*}
and so the inner product of \eqref{eq:eff} with $\G_1$ gives rise to
\begin{equation*}
  0 =  \left(\alpha + i\,\langle\G_1,\,\G_2\rangle\,\delta\right)\,\langle\G_1,\,\G_2\rangle\,.
\end{equation*}
Thus, for mutually non-local charge vectors, $\alpha=\delta=0$. Similarly,
computing the inner product with $\G_2$ and using the antisymmetry
properties of it, we get
\begin{equation*}
  0 =  \left(\beta + i\,\langle\G_1,\,\G_2\rangle\,\gamma\right)\,\langle\G_1,\,\G_2\rangle\,.
\end{equation*}
Once again, for mutually non-local charge vectors, we must conclude
$\beta=\gamma=0$. All in all, we learn that there is no $Z_{\star}$
trivially satisfying \eqref{posreq}, being consistent with charge
conservation and having a 1/8 BPS constituent with vanishing quartic
invariant. Any other choice of $Z_{\star}$ would give rise to a further constraint on the moduli.

We conclude that any pair of
charge vectors $\{\G_1,\,\G_2\}$ with $\G_1$ being 1/8 BPS with
$\Delta_1=0$, consistent with supersymmetry and charge conservation
will have some extra moduli dependent condition ensuring the
positivity of $\Sigma^2$ close to the 1/8 BPS center $\vec{x}_1$ and
necessarily increasing the co-dimension of its wall of marginal
stability.

\section{Discussion}

We studied the gravitational realization of supersymmetric D0-D6
bound states in the STU model. In the large volume limit, we
determined all supersymmetric regular two-centered configurations
consistent with the composites of the system existing in regions of
moduli space bounded by a wall of marginal stability of co-dimension
one. The possible constituents states of the system are
\begin{equation}
\begin{split}\label{Fstate}
\Gamma_1&=(p^0,[-p];0,0)~,\quad \Gamma_2=(0,[p];0,q_0)~,\\
\Gamma_1&=(p^0,0;[q],0)~,\quad \Gamma_2=(0,0;[-q],q_0)~.
\end{split}
\end{equation}
The domain in moduli space where the bound state exists is described
by \eqref{wall4} and \eqref{phase4}. The shape of these walls is
analogous to those first found in \cite{Sen:2007vb,Cheng:2007ch}. At
this level, $p$ and $q$ are only constrained by our discussion in
section \ref{sec:existence}. After imposing charge quantization on
the vectors \eqref{Fstate}, {\it i.e.} discrete U-duality group, the
final states \eqref{Fstate} will be further reduced.

We have explicitly seen how global requirements of regularity imposed
additional constraints on the existence of the state, besides the
more kinematical (or algebraic) characterization of the charge
vectors and their central charges. In other words, the local
conditions from supersymmetry and regularity of the central charge
are necessary but {\it not} sufficient to provide a well-defined
supergravity configuration.

An intuitive explanation for this fact is that all allowed
constituents for the system have vanishing quartic invariant. As
such, they are singular solutions to the attractor equations.
Whenever each of these builds a bound state, the dominant
contribution to the behavior of the metric close to the center where
such charge sits is no longer determined purely in terms of the
charges. In addition it also depends on the moduli and the phase of
the overall central charge of the bound state, which means that
positivity of $\Sigma$ in that location is already a non-trivial
requirement. Indeed, we have seen that only for certain constituents
such behavior is guaranteed to be positive whenever we are in the
appropriate side of the wall of marginal stability, {\it i.e.}
whenever the bound state was algebraically supposed to exist.
Interestingly, whenever this requirement is fulfilled, we can also
prove that the solution is free of CTCs. This observation will also
be relevant for any multi-center configuration built of constituents
having vanishing quartic invariants.

It would be interesting to extend our results to the full ${\cal
N}=8$ theory. The additional moduli of $E_7$  will likely impose
additional constraints on the phases of the central charge
\cite{Sen:2008ta}. It is also clearly meaningful to apply our
techniques to more general situations involving polar states with
$\Delta>0$ and attempting to relate them to the attractor flow
conjecture and entropy enigma presented in \cite{Denef:2007vg}.

\newpage

\section*{Acknowledgements}
We thank Miranda Cheng, Jim T. Liu and Sameer Murthy for discussions. The work of
AC is supported in part by DoE under the grant DE-FG02-95ER40899.
The work of JS was partially supported by the Engineering and Physical Sciences Research
Council [grant number EP/G007985/1]. This research was supported in part by the National
Science Foundation under Grant No. NSF PHY05-51164.
AC would like to thank the organizers of the ``Monsoon Workshop on String theory'' at TIFR and ``Spring School on Super String Theory
and Related Topics'' at ICTP for hospitality during stages of this work. JS would like to thank the organisers of the KITP programme "Fundamentals of String Theory" for hospitality during the
final stages of this project.


\appendix

\section{Multi-centered solutions in the STU-model} \label{sec:sugra}

When $g_s|\Gamma|\gg 1$, we expect the supergravity approximation
to provide a reliable description of any state in the
theory. As reviewed in \cite{Denef:2007vg}, the exact description
will depend on the existence of the state in moduli space. More
precisely, if the central charge $Z(\Gamma)$ corresponding to a given
charge vector $\Gamma$ vanishes at a regular point in moduli space,
the single centered supergravity solution will {\it not} exist. This
is indeed the case for $\Gamma_{\text{D0-D6}}$. In such situations,
these states can be realized in terms of multi-centered supergravity
configurations, which are stationary but non-static.

In the following, we will present a very brief review of the relevant
multi-centered black hole solutions constructed in
\cite{Denef:2000nb,LopesCardoso:2000qm,Bates:2003vx}. A more recent
discussion can be found in \cite{Denef:2007vg, deBoer:2008fk}. The
four-dimensional metric, gauge fields and moduli are given by
\begin{equation}
\begin{split}\label{multicenter}
 ds^2 & =  -\frac{1}{\S}(dt+\omega)^2+\S\, ds_{\mathbb{R}^3}^2\,,\\
 \CA^0 & =  \frac{\partial \log \S}{\partial H_0}\left(dt +\omega\right)+\omega_0\,,\\
 \CA^A & =  \frac{\partial \log \S}{\partial H_A}\left(dt +\omega\right)+\CA_d^A\,,\\
 z^A&=\frac{H^A-i\frac{\partial \S}{\partial H_A}}{H^0+i\frac{\partial \S}{\partial H_0}},
\end{split}
\end{equation}
where $H=\left(H^\Lambda;\,H_\Lambda\right)$ is a set of harmonic
functions in $\RR^3$ which encodes the location of charges at each
center. Explicitly we have
\begin{subequations}\label{App:harmonic}
\begin{align}
& H^\Lambda=\sum_{i=1}^N {p^\Lambda_i\over |\vec{x}-\vec{x}_i|}+h^\Lambda~,\label{harmonic1}
\\
&H_\Lambda=\sum_{i=1}^N {q^i_\Lambda\over |\vec{x}-\vec{x}_i|}+h_\Lambda~,\label{harmonic2}
\end{align}
\end{subequations}
with $N$ the total number of centers. A priori, it is allowed to have
an arbitrary number of centers $\vec{x}_i$ carrying charges $\Gamma_i
= \left(p^\Lambda_i\,;q^i_\Lambda\right)$. The vector $h =
\left(h^\Lambda;h_\Lambda\right)$ stands for constants characterizing
the asymptotic value of all the harmonic functions. More explicitly,
it is given in terms of the asymptotic moduli and the phase $\alpha$
of the total central charge by
\begin{equation}
\begin{split}\label{moduli}
  h^0 &= -2e^{K/2}\sin\alpha\,, \\
  h^A &=  2e^{K/2}\left(\cos\alpha\,\text{Im}(z^A)-\sin\alpha\,\text{Re}(z^A)\right)\,,\\
  h_A &= 2e^{K/2}\left(\cos\alpha\,\text{Im}\left(\frac{1}{2}s_{ABC}z^Bz^C\right)-\sin\alpha\,\text{Re}\left(\frac{1}{2}s_{ABC}z^Bz^C\right)\right)\,, \\
  h_0 &= 2e^{K/2}\left(\cos\alpha\,\text{Im}(z^1z^2z^3)-\sin\alpha\,\text{Re}(z^1z^2z^3)\right)\,,
\end{split}
\end{equation}
where $K$ is defined by \eqref{Kahler}. It is understood that in
\eqref{moduli} all moduli dependence is evaluated at spatial
infinity, {\it i.e.} $z^A=z^A_\infty$.

Restricting the discussion to Type IIA compactified on a 6-torus (in
its STU-truncation), the factor $\Sigma$ in \eqref{multicenter} is
uniquely given by
\begin{align}\label{App:sigma}
  \Sigma^2(H) = &-(H_\Lambda H^\Lambda)^2+4\left(H^1H_1H^2H_2+H^1H_1H^3H_3+H^2H_2H^3H_3\right)\nonumber\\ &-4H^0H_1H_2H_3-4H_0H^1H^2H^3~.
\end{align}
Notice  $\Sigma^2(H)$ is nothing but the quartic invariant
\eqref{quartic} in which all charges
$\Gamma=\left(p^\Lambda;\,q_\Lambda\right)$ have been replaced by the
harmonic functions $H=\left(H^\Lambda;\,H_\Lambda\right)$.

The off diagonal metric components can be found explicitly by solving
\begin{equation} \label{omegaeom}
 \star d\omega = \langle dH,H\rangle \, ,
\end{equation}
where $\star$ is the Hodge dual on flat $\mathbb{R}^3$. The Dirac
parts $\CA_d^A$, $\omega_0$ of the vector potentials can be obtained
from
\begin{equation}
 d\omega_0 =  \star dH^0 \,, \quad
 d\CA_d^A = \star dH^A \,.
\end{equation}

Regularity of the solution requires $N-1$ independent consistency
conditions on the relative positions of the $N$ centers, reflecting
the fact that these configurations really are interacting and one
can't move the centers around freely. These conditions arise from
requiring integrability of \eqref{omegaeom}
\begin{equation}
  \langle H,\G_i\rangle|_{x=x_i}=0\,,
 \label{consistency}
\end{equation}
or written out more explicitly
\begin{equation}  \sum_{b \neq a}
   \frac{\langle
    \G_a,\G_b\rangle}{r_{ab}}=\langle h,\G_a\rangle\,, \quad \text{with} \quad \langle \G_i,\,\G_j\rangle = -p_i^0q_0^j + p_i^Aq_A^j - q_A^ip^A_j + q_0^ip_j^0\,.
  \label{consistency2}
\end{equation}
where $r_{ab}=|x_a-x_b|$. Consequently, the equilibrium distances
between the different centers  depend on the asymptotic values of the
scalar fields and on the charges at each center.

A crucial property of these multi-centered solutions is that they
carry intrinsic angular momentum due to rotations on $\RR^3$, which
equals to
\begin{equation}
  \vec{J} = \sum_{i<j} \frac{1}{2}\,\langle \Gamma_i,\,\Gamma_j\rangle\,\frac{\vec{x}_i-\vec{x}_j}{|\vec{x}_i-\vec{x}_j|}~.
\end{equation}
Due to the off-diagonal terms in the metric sourcing this angular
momentum, there are further requirements this set of configurations
have to satisfy to prevent the existence of closed timelike curves
(CTC). These are guaranteed to be absent if
\begin{align}\label{CTC}
\Sigma^2 > \omega_i\omega^i~,
\end{align}
a condition that has to be satisfied everywhere, and not just point
wise \cite{Berglund:2005vb,Bena:2005va}.\footnote{This condition may
not be satisfied and the configuration still be free of these causal
pathologies, i.e. this condition is sufficient, but not necessary.}

Assuming a given charge vector $\Gamma_i$ solves the attractor
equations, the behavior of the multi-centered solution close to the
center $\vec{x}_i$ is fully determined by the charges in $\Gamma_i$,
due to the attractor mechanism. In particular, $\Sigma$ is a function
of the entropy of the pole, {\it i.e.} the quartic invariant
evaluated at that center. But depending on the amount of
supersymmetry preserved by the state associated with $\Gamma_i$, such
entropy might vanish. Under these circumstances, the order of the
pole changes. More importantly, the value of the pole will {\it no}
longer be determined by the attractor mechanism. For now, we are just
interested in matching the order of the pole with the amount of
supersymmetry preserved by the state.

According to the discussion in \cite{4dmicro}, the
prescription is that by looking at the scaling of $\Sigma^2(H)$ with
respect to the distance to the center $\rho=|x-x_i|\to 0$, one finds
:
\begin{equation}\label{susy}
\begin{array}{ccc}
{1/8}~{\rm BPS~,}& ~\Delta>0~, ~&\Sigma^2\propto \rho^{-4}\\
{1/8}~{\rm BPS~,}& ~\Delta=0~, ~&\Sigma^2\propto \rho^{-3}\\
{1/4}~{\rm BPS~,}& ~\Delta=0~, ~\partial\Delta=0~, &\Sigma^2\propto \rho^{-2}\\
{1/2}~{\rm BPS~,}& ~\Delta=0~, ~\partial\Delta=0~,~\partial^2|_{Adj}\Delta=0~, &\Sigma^2\propto \rho^{-1}
\end{array}
\end{equation}
where the symbol $\partial$ denotes derivatives with respect to the
charges $p^\Lambda$ and $q_\Lambda$.

\section{Algebraic description of 1/4 and 1/2 BPS states}
\label{sec:proof}

Both 1/4 and 1/2 BPS states have vanishing quartic invariant and
vanishing $\partial\Delta/\partial q_\Lambda =
\partial\Delta/\partial p^\Lambda=0$. The latter set of conditions is
:

\begin{eqnarray}
{\partial \Delta\over\partial q_0}&=& 2p^0(p^\Lambda q_\Lambda)-4p^1p^2p^3=0\,, \label{eq:pr1} \\
{\partial \Delta\over\partial p^0}&=& 2q_0(p^\Lambda q_\Lambda)-4q_1q_2q_3=0\,, \label{eq:pr2} \\
{\partial \Delta\over\partial q_A}&=& -2p^A(p^\Lambda q_\Lambda)+4p^A\sum_{B\neq A}p^Bq_B-2p^0s_{ABC}q_Bq_C=0\,, \label{eq:pr3} \\
{\partial\Delta\over\partial p^A}&=& -2q_A(p^\Lambda q_\Lambda)+4q_A\sum_{B\neq A}p^Aq_B-2q^0s_{ABC}p^Bp^C=0\,. \label{eq:pr4}
\end{eqnarray}

Let us assume $p^0,\,q_0\neq 0$. Using \eqref{eq:pr1} and \eqref{eq:pr2}, we learn that
\begin{equation*}
  p^0q_1q_2q_3 = q_0p^1p^2p^3 \quad \Leftrightarrow \quad
  p^\Lambda q_\Lambda = \frac{2}{q_0}\,q_1q_2q_3 = \frac{2}{p^0}\,p^1p^2p^3\,.
\end{equation*}
Multiplying \eqref{eq:pr3} with $q_A$ (without summing over the index
$A$) we obtain :
\begin{equation}
  -2p^Aq_A\,(p^\Lambda q_\Lambda) - 4p^0q_1q_2q_3 + 4p^Aq_A\sum_{B\neq A} p^Bq_B = 0\,.
 \label{eq:ap1}
\end{equation}
Using the identities :
\begin{eqnarray*}
  4p^Aq_A\sum_{B\neq A} p^Bq_B &=& -4(p^Aq_A)^2 +4p^Aq_Ap^0q_0 + 4p^Aq_A(p^\Lambda q_\Lambda)\,, \\
  -2p^Aq_A\,(p^\Lambda q_\Lambda) - 4p^0q_1q_2q_3 &=& -2(p^\Lambda q_\Lambda)\left(p^0q_0+p^Aq_A\right)\,,
\end{eqnarray*}
we can write \eqref{eq:ap1} as
\begin{equation*}
  2\left(\left(p^\Lambda q_\Lambda\right)-2p^Aq_A\right)\left(p^Aq_A-p^0q_0\right) = 0\,,
\end{equation*}
where there is still no summation over the index $A$. It is
convenient to introduce the auxiliary variables $x_0=p^0q_0$ and
$x_A=p^Aq_A$ for $A=1,2,3$ to solve this equation :
\begin{equation*}
  \left(-x_0+x_1+x_2+x_3-2x_A\right)\left(x_A-x_0\right) = 0\,.
\end{equation*}
In terms of these variables, it is easy to find  the general solution :
\begin{equation*}
  x_A = x_0\,, \quad \quad x_B=x_C \quad \quad A\neq B \neq C
\end{equation*}
up to permutations in the three tori, i.e. $A\leftrightarrow B
\leftrightarrow C$. It is the above fact that allows us to write the
charge vector in terms of eight parameters
$\{\alpha_{1,2},\beta_{1,2}\}$ and $\{P^0,P^2,P^3,Q_0\}$ :
\begin{equation}
  \Gamma = \left(\beta_1P^0,\beta_2P^0,\alpha_1P^2,\alpha_1P^3,\beta_1Q_0,\alpha_2P^3,\alpha_2P^2,\beta_2Q_0\right)
 \label{eq:fvector}
\end{equation}
Inserting this expression in our initial set of equations \eqref{eq:pr1}-\eqref{eq:pr4}, we obtain :
\begin{equation}
\begin{array}{cc}\label{eq:fcons}
P^0P^2P^3\,\alpha_{1,2}\,(\beta_1\alpha_2-\beta_2\alpha_1)&=0\,,  \\
Q_0P^2P^3\,\alpha_{1,2}\,(\beta_1\alpha_2-\beta_2\alpha_1)&=0\,, \\
Q_0P^0P^{2,3}\,\beta_{1,2}\,(\beta_1\alpha_2-\beta_2\alpha_1)&=0\,.
\end{array}
\end{equation}
whereas the vanishing of the quartic invariant $\Delta$ requires :
\begin{equation*}
  \Delta = -4\left(\beta_1\alpha_2-\beta_2\alpha_1\right)^2\,P^0P^2P^3Q_0\,.
\end{equation*}
This latter constraint is not independent, since whenever all the first derivatives of $\Delta$ vanish,
the quartic invariant itself also does.

Thus, for non-vanishing parameters, the solution will always be given
by $\beta_1\alpha_2=\beta_2\alpha_1$. But we can still satisfy
\eqref{eq:fcons} by setting a combination of coefficients
$(\alpha_{1,2},\beta_{1,2})$ and/or charges $(P^0,\,P^2,\,P^3,\,Q_0)$
to vanish.

The previous derivation assumed that both $(p^0,q_0)$ were not
vanishing.\footnote{Strictly speaking, when multiplying our initial
equations by $q_A$ and $p^A$ we were also assuming all charges were
generically non-vanishing.} It is easy to extend the analysis when
either of them vanishes.

\paragraph{The $q_0=0$ branch.} Equation \eqref{eq:pr2} implies the product $q_1q_2q_3$ vanishes.
Let us pick one of them to vanish, i.e. $q_A=0$ (for some A=1,2,3),
having $q_B,\,q_C\neq 0$ for $B\neq C\neq A$. In this situation,
$(p^\Lambda q_\Lambda) = x_B + x_C$, where $x_B$'s were defined as
above. The non-trivial equations to solve become :
\begin{eqnarray*}
  \frac{\partial\Delta}{\partial q_B} &=& 2p^B\left(x_C-x_B\right) = 0\,, \quad B\neq C,\, B,C\neq A \\
  \frac{\partial\Delta}{\partial p^B} &=& 2q_B\left(x_C-x_B\right) = 0\,, \quad B\neq C,\, B,C\neq A \\
  \frac{\partial\Delta}{\partial q_A} &=& 2p^A(p^\Lambda q_\Lambda) - 4p^0q_Cq_B = 0\,, \\
  \frac{\partial\Delta}{\partial q_0} &=& 4p^1p^2p^3-2p^0(p^\Lambda q_\Lambda) = 0\,.
\end{eqnarray*}

If all charges appearing above are generically non-zero, the solution is given by :
\begin{equation*}
  x_B = x_C \quad \quad \text{and} \quad \quad p^0 = \frac{p^A\,p^B}{q_C}\,,
\end{equation*}
which is the particular case $Q_0=0$ in the charge vector \eqref{eq:fvector}.

If we do not impose $x_B=x_C$, we are forced to allow charges to
vanish, and we always end up satisfying $x_B=x_C=0$. The most general
set of solutions in this category are summarised by
\begin{equation*}
  \left(p^0,[p^A,p^B];[q_C],0\right) \quad \quad \text{and} \quad\quad \left(0,[p^A];[q_B,q_C],0\right)
\end{equation*}
which do still belong to the class described by \eqref{eq:fvector},
without necessarily satisfying the condition
$\beta_1\alpha_2=\beta_2\alpha_1$.

\paragraph{The $p^0=0$ branch.} The analysis of this branch is completely analogous to the one above.
In this case, one of the $p^A$ charges has to vanish because of
\eqref{eq:pr1}. If all remaining charges are non-vanishing, we again
have $x_C=x_B$, with $q_0=q_Aq_B/p^C$. If extra charges are allowed
to vanish, all solutions are included in either of the following two
sets :
\begin{equation*}
  \left(0,[p^A,p^B];[q_C],0\right) \quad \quad \text{and} \quad\quad \left(0,[p^A];[q_B,q_C],q_0\right)
\end{equation*}
which do still belong to the class described by \eqref{eq:fvector},
without necessarily satisfying the condition
$\beta_1\alpha_2=\beta_2\alpha_1$.

\paragraph{Conclusion.} The analysis presented above proves that any 1/4 or 1/2 BPS state has a charge vector of the form \eqref{eq:fvector} :
\begin{equation*}
\Gamma = \left(\beta_1P^0,\beta_2P^0,\alpha_1P^2,\alpha_1P^3,\beta_1Q_0,\alpha_2P^3,\alpha_2P^2,\beta_2Q_0\right)
\end{equation*}
where either $\beta_1\alpha_2=\beta_2\alpha_1$, or whenever
$\beta_1\alpha_2\neq\beta_2\alpha_1$, there are enough vanishing
coefficients and/or charges so that \eqref{eq:fcons} are still
satisfied.

\section{Moduli identities}
\label{sec:moduli}

In this appendix we gather some useful expression relating constant
asymptotic value of the harmonic functions $(h^\Lambda,h_\Lambda)$
and the moduli $z^A$ that we used in section \ref{sec:regsugra}. The
total charge of the system is $\Gamma_{\rm D0-D6}$ and the central
charge is
\begin{equation}
Z_{\rm D0-D6}=e^{K/2}(p^0z^1z^2z^3-q_0)=|Z_{\rm D0-D6}|e^{i\alpha}
\end{equation}
Starting from the definitions \eqref{moduli}, we have
%

%
\begin{equation}\label{app:modh1}
\begin{split}
h^1&=-{2e^{K/2}\over |Z_{\rm D0-D6}|}\left(q_0J^1+p^0|z^1|^2{\rm Im}(z^2z^3)\right)~,\\
h_1&=-{2e^{K/2}\over |Z_{\rm D0-D6}|}\left(q_0{\rm Im}(z^2z^3)+p^0J^1|z^2z^3|^2\right)~,
\end{split}
\end{equation}
and similarly expressions for $h^{2,3}$ and $h_{2,3}$, where
$z^A=B^A+iJ^A$. In the function $\Sigma^2$ for type {\bf I} bound
states the following combinations appear

\begin{equation}\label{app:mod4a}
\begin{split}
h_2h_3-h^1h_0&=4e^{K}|z^1|^2J^2J^3~,\\
h^2h^3-h_1h^0&=4e^{K}J^2J^3~,\\
h_2h^2+h_3h^3-h_1h^1-h_0h^0&=8e^{K}B^1J^2J^3~.
\end{split}
\end{equation}
Linear combinations of these terms in \eqref{poleI} simplify to
$(4e^{K}J^2J^3|az^1-c|^2)$ and $(4e^{K}J^2J^3|dz^1-b|^2)$. Finally,
other useful identities are
\begin{equation}\label{app:modhh}
\begin{split}
h^1h_1-h_0h^0&=4e^{K}J^1 {\rm Im}(z^2z^3)~,\\
h^2h_2-h_0h^0&=4e^{K}J^2 {\rm Im}(z^1z^3)~,\\
h^3h_3-h_0h^0&=4e^{K}J^3 {\rm Im}(z^1z^2)~.
\end{split}
\end{equation}

\bibliographystyle{utphys}

\bibliography{CS08}

\end{document}